\documentclass{egpubl}
\usepackage{eg2024}

\ConferenceSubmission   
\usepackage[T1]{fontenc}
\usepackage{dfadobe}  
\usepackage{wrapfig}
\usepackage{graphicx} 
\usepackage{units}
\usepackage{amsmath}
\usepackage{arydshln}
\newcommand{\new}[1]{{\color{black}#1}}
\usepackage{cite}  
\BibtexOrBiblatex
\electronicVersion
\PrintedOrElectronic

\usepackage{egweblnk}
\usepackage{amssymb} 
\usepackage[ruled]{algorithm2e} 

\SetAlFnt{\small}
\SetAlCapFnt{\small}
\SetAlCapNameFnt{\small}
\SetAlCapHSkip{0pt}

\usepackage{hyperref}

\title[\textsc{FlameForge}]%
      {\textsc{FlameForge}: Combustion of Generalized Wooden Structures}

\author[1018]{D. Liu, J. Klein, F. Rist (KAUST); W. Pałubicki (AMU); S. Pirk (CAU); D. L. Michels (KAUST)}
\usepackage{color}
\usepackage{xcolor}

\begin{document}

\teaser{
\includegraphics[height=15em]{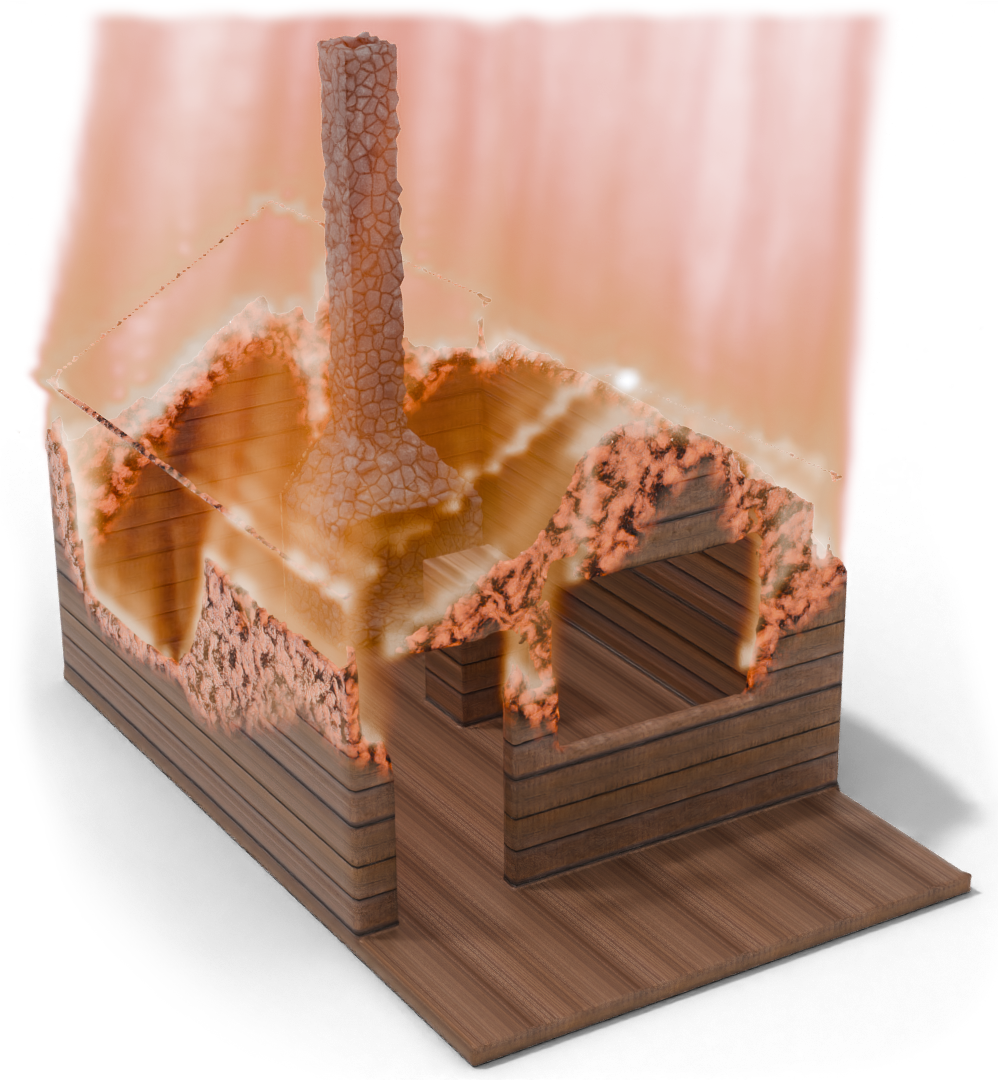} \hfill{}
\includegraphics[height=15em]{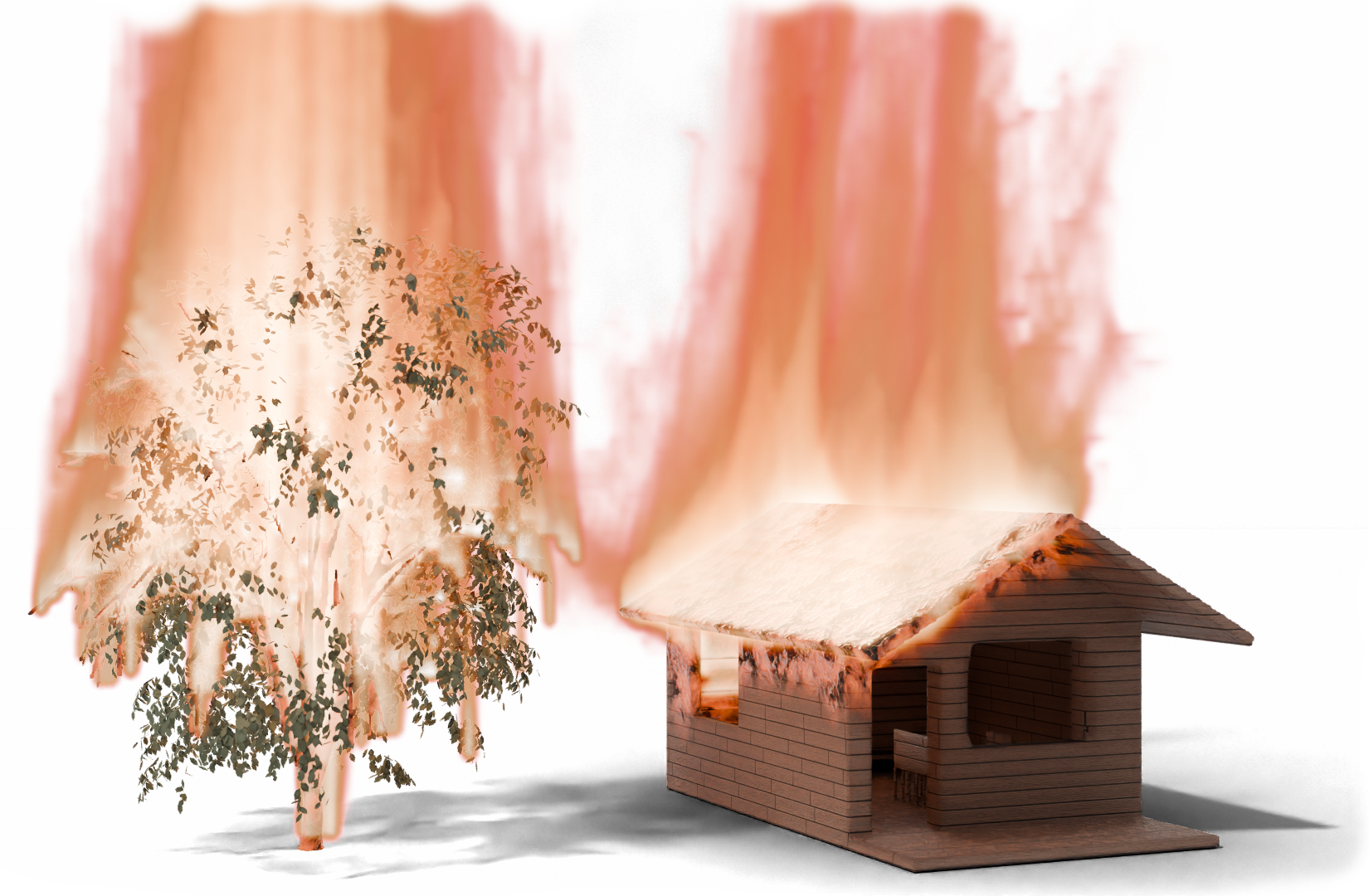}\hfill{}
\includegraphics[height=15em]{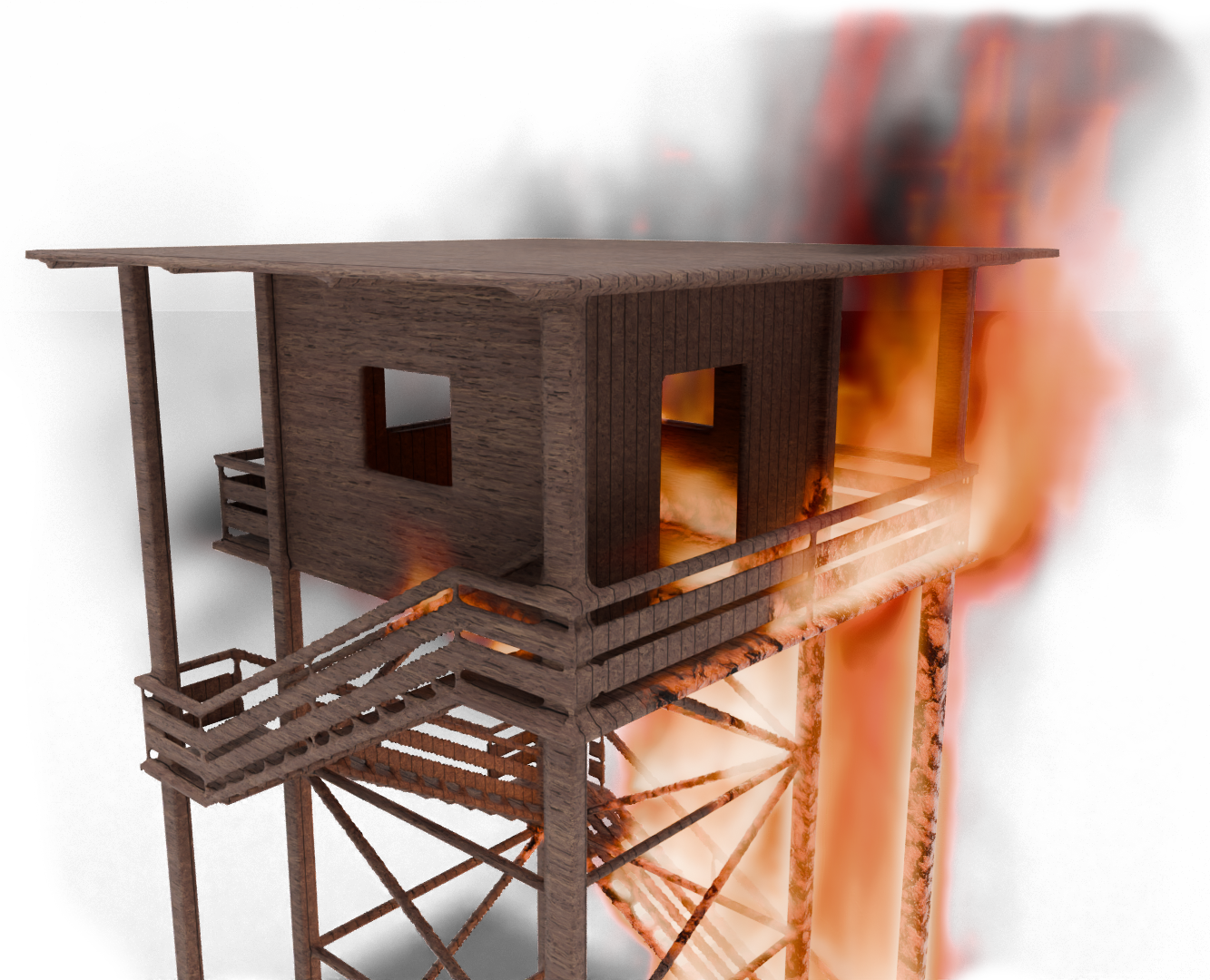}
\caption{Illustrations of different scenes showcasing the capabilities of our \textsc{FlameForge} simulator. Left: A house has been set on fire which is composed of two types of wood (walls/roof), acrylic windows, and a chimney made out of stone. Middle: The fire of a burning tree further spreads to the roof of a wooden house. Right: A watchtower has been ignited at the bottom causing both horizontal as well as strong vertical fire spread.}
\label{fig:teaser}
}

\maketitle
\begin{abstract}
\stepcounter{footnote}

We propose a unified volumetric combustion simulator that supports general wooden structures capturing the multi-phase combustion of charring materials. Complex geometric structures can conveniently be represented in a voxel grid for the effective evaluation of volumetric effects. In addition, a signed distance field is introduced to efficiently query the surface information required to compute the insulating effect caused by the char layer. Non-charring materials such as acrylic glass or non-combustible materials such as stone can also be modeled in the simulator. Adaptive data structures are utilized to enable memory-efficient computations within our multiresolution approach. The simulator is qualitatively validated by showcasing the numerical simulation of a variety of scenes covering different kinds of structural configurations and materials. Two-way coupling of our combustion simulator and position-based dynamics is demonstrated capturing characteristic mechanical deformations caused by the combustion process. The volumetric combustion process of wooden structures is further quantitatively assessed by comparing our simulated results to sub-surface measurements of a real-world combustion experiment.\\
Supplementary Video: \url{https://youtu.be/3q6q56NU1Vo}


\end{abstract}
\section{Introduction}

Wood is a venerable and sustainable building material that has been a cornerstone of construction throughout human history.
Structures built out of wood play a crucial role for construction, offering a unique blend of functionality, sustainability, and aesthetic appeal.

As a renewable resource, wood also contributes to environmentally conscious construction practices, promoting a reduced carbon footprint compared to other building materials. According to Himes and Busby~\shortcite{HIMES2020100030}, ``substituting conventional building materials for mass timber reduces construction phase emissions by 69\%, an average reduction of 216\,kg\,CO$_2$e/m$^2$ of floor area. [$...$] Scaling-up low-carbon construction, assuming mass timber is substituted for conventional building materials in half of expected new urban construction, could provide as much as 9\% of global emissions reduction needed to meet 2030 targets for keeping global warming below 1.5$^\circ$C.''

Its innate strength-to-weight ratio makes wood an ideal choice for a variety of structural applications, from residential homes and commercial buildings to bridges and pavilions.

Understanding the combustion behavior of wooden structures assumes paramount importance due to its widespread use.
The combustion of wooden structures not only poses challenges for fire safety but also holds significant implications for environmental impact and sustainable design practices.
In this paper, we aim for the numerical simulation of the combustion process of general wooden structures. 
Our work is motivated by the desire to gain deeper insights into their mechanical performance, stability, and overall structural behavior.

Combining mathematical modeling and numerical simulation with effective handling of complex underlying geometric structures has always been at the core expertise of the animation and simulation community within the field of computer graphics.
The simulation of fire-related phenomena have been addressed by the community~\cite{nielsen2019physics,10.1145/3587421.3595406} as fire emerges as an elemental force, lending a visceral and captivating authenticity to computer graphics scenes in films and games.
Among others, the combustion of wooden structures has been addressed by Pirk et al.~\shortcite{pirk2017interactive} focusing solely on wood combustion for botanical tree models by exploiting their specific geometric properties. In contrast, our work aims for the simulation of the combustion process of general wooden structures as shown in Figure~\ref{fig:teaser}.

In this regard, our specific contributions are as follows:
We (1) propose a unified combustion framework that supports arbitrary geometries and complex materials focusing on the multi-phase combustion of charring materials such as wood, but also non-charring materials such as acrylic glass or non-combustible materials such as stone;
we (2) propose a representation of linked voxel grids and signed distance fields to efficiently evaluate volumetric and surface effects;
we (3) employ adaptive data structures to effectively handle sparse areas in a memory-efficient manner;
we (4) qualitatively evaluate our approach on a variety of scenes covering different kinds of geometries and materials, environmental conditions such as wind, and the spread of fire along the same or across different objects;
we (5) demonstrate two-way coupling with position-based dynamics to capture characteristic mechanical deformations caused by the combustion process;
we (6) quantitatively assess the accuracy of our fully volumetric combustion model including the insulating effect of the char layer by comparing simulated results to sub-surface measurements taken from our real-world combustion experiment.
\section{Related Work}

Computational fluid dynamics is an active field of research in different scientific communities, including computer graphics and visual computing~\cite{10.1145/311535.311548,Huang:2021:VastOcean,Shao:2022:Multigrid}. The domain of combustion within computational fluid dynamics has been the target of intensive study, with foundational texts including Kroos and Potter~\shortcite{kroos2014thermodynamics}, Merci and Beji~\shortcite{2016fluid}, Versteg and Malalasekera~\shortcite{versteeg2007introduction}, and Peters~\shortcite{peters2000turbulent} offering essential insights. For an in-depth understanding of fire simulation techniques specific to computer graphics, Huang et al.'s ~\shortcite{Huang2014} survey is as an informative reference. 

Several authors have explored diverse techniques in fire simulation, with a notable emphasis on combining visual modeling with procedural and statistical methods. This approach has been employed to produce various effects, such as flames~\cite{10.1145/566654.566644}, providing benefits like reduced computation time and enhanced user control. However, these methods face challenges in capturing the full spectrum of realism that physics-based techniques can offer, a notion echoed in earlier works of Hong et al.~\shortcite{10.1007/s00371-009-0403-8} and Kim et al.~\shortcite{10.1007/s00371-016-1267-3}, who have highlighted the complexities in controlling physics-based fire simulations.

In the direction of physics-based modeling, Stam and Fiume~\shortcite{10.1145/218380.218430} presented a significant contribution by proposing a fire model that integrates finite-rate chemical kinetics, convection, conduction, radiative cooling, and a ray casting approximation for radiative heating. This approach was further expanded by Ihm et al.~\shortcite{10.2312:SCA:SCA04:203-212} and Kang et al.~\shortcite{10.2312:SCA:SCA07:199-208}, who combined chemical kinetics with heat transfer to simulate fire and explosions. These studies provide a contrast to methodologies that assume infinitely fast chemistry, a simplification often necessary due to the short time-scales of chemical reactions compared to the requirements of computer graphics. This trade-off, while facilitating certain aspects of simulation, presents its own set of challenges and opportunities in the pursuit of realistic flame representation. Nielsen et al.~\shortcite{nielsen2022physics} introduced a sophisticated physics-based combustion simulation method for computer graphics, enhancing realism in flame, temperature, and soot distribution modeling. Their approach integrates comprehensive thermodynamic models of real-world fuels and advanced heat transfer methods, enabling the accurate simulation of diverse deflagration phenomena.

Our work is closest to Pirk et al.~\shortcite{pirk2017interactive} who developed a method for simulating tree combustion using a structure of connected particles and a polygonal mesh, enabling realistic dynamics in branch motion and fire propagation, validated against real wood samples. Expanding on this, H\"adrich et al.~\shortcite{hadrich2021fire} introduced a large-scale wildfire simulation using detailed 3D tree models, incorporating advanced mathematical formulations for plant combustion and fire spread, aimed at realistically modeling the impact of various environmental and preventive measures on wildfires. This work has been later extended by Kokosza et al.~\shortcite{kokosza2024scintilla}. Melek and Keyser~\shortcite{1167889} developed a linear combustion model, distinctly representing fuel, oxygen, and combustion products. Their model included convection and conduction in heat transfer, but their approach to radiation, based on a diffusion process, did not fully capture the ability of radiation to propagate through space without heating it.

Nguyen et al.~\shortcite{nguyen2002physically} introduced the thin flame model to computer graphics. Their physics-based model effectively captured low-speed deflagrations like laminar and turbulent flames from premixed fuel combustion. The flames were modeled as an inner closed surface blue core, where the chemical reaction of fuel and oxidizer occurs. This model conserved mass and momentum across the flame front, reflecting changes in velocity and density at the flame. 
Feldman et al.~\shortcite{feldman2003} introduced a method for simulating suspended particle explosions using a combination of Eulerian and Lagrangian components, interacting through drag forces and heat transfer. The Eulerian approach simulates air, incorporating convection, conduction, and approximate radiative heating similar to Melek and Keyser~\shortcite{1167889}, as well as radiative cooling via the Stefan-Boltzmann law. The Lagrangian particles in their model carry fuel and soot, employing linear combustion and soot production models.

Losasso et al.~\shortcite{10.1109/TVCG.2006.51} expanded upon the model of Nguyen et al.~\shortcite{nguyen2002physically} by including the erosion and conversion of solids into gaseous fuel, accounting for the volumetric expansion in this conversion process. Similarly, Hong et al.~\shortcite{10.1145/1276377.1276436} added a physics-based model for wrinkled flames to Nguyen et al.'s~\shortcite{nguyen2002physically} method and introduced a jump condition for a more accurate temperature profile across the flame front. Additionally, Stomakhin et al.~\shortcite{Stomakhin:2014:AMP:2601097.2601176} introduce a point-based technique for simulating the melting and solidifying of materials by modeling heat transfer, capturing varying thermodynamic properties and altering mechanical properties, albeit without support for interactive modeling and real-time heat diffusion.


Kwatra et al.~\shortcite{10.5555/1921427.1921458} proposed a fuel combustion model consistent with earlier works \cite{horvath2009directable,feldman2003}, where fuel combustion and soot production are modeled through a linear ODE. This approach to fuel combustion has been a consistent theme in the field, contributing to the broader understanding of the dynamics involved. In the broader context of combustion simulation, which is a staple in production for simulating fire and deflagrations at various scales, significant advancements have been made. These include fire simulation on GPUs advanced particle workflows~\cite{horvath2009directable}.

For smoke, fire, and fluid simulation, efforts have been made to enhance the temporal accuracy of Stam’s~\shortcite{Stam:1999:SF:311535.311548} semi-Lagrangian stable fluids approach while maintaining its stability for large time steps. A noteworthy advancement is the second-order accurate advection-reflection solver by Narain et al.~\shortcite{10.1145/3340257}. This solver combines two advection steps with two projections per time step, aiming to reduce energy dissipation due to projection and improve the tracking of vortical features in simulations. Comparatively, it shows significant advantages over the second-order accurate BDF2-related scheme by Xiu and Karniadakis~\shortcite{xiu2001}. Although the latter is more cost-effective per time step, involving two advection steps but only one projection, it tends to yield less accurate results even when time steps are adjusted for equal compute time.

Finally, we would like to mention that next to phenomena related to fire, there is also an emerging body of work dedicated to simulating complex natural phenomena and other natural disasters. For example, recent efforts have successfully simulated the dynamics of hurricanes and tornadoes~\cite{AmadorHerrera:2024:Cyclogenesis} and innovative techniques have been developed to model atmospheric conditions and cloud dynamics, which are crucial for understanding the broader environmental context~\cite{Haedrich:2020:Stormscapes,AmadorHerrera:2021:Weatherscapes,palubicki2022ecoclimates}.


\section{Methodology}
\label{sec:Methodology}

\begin{figure}
\centering
\includegraphics[width=0.85\columnwidth]{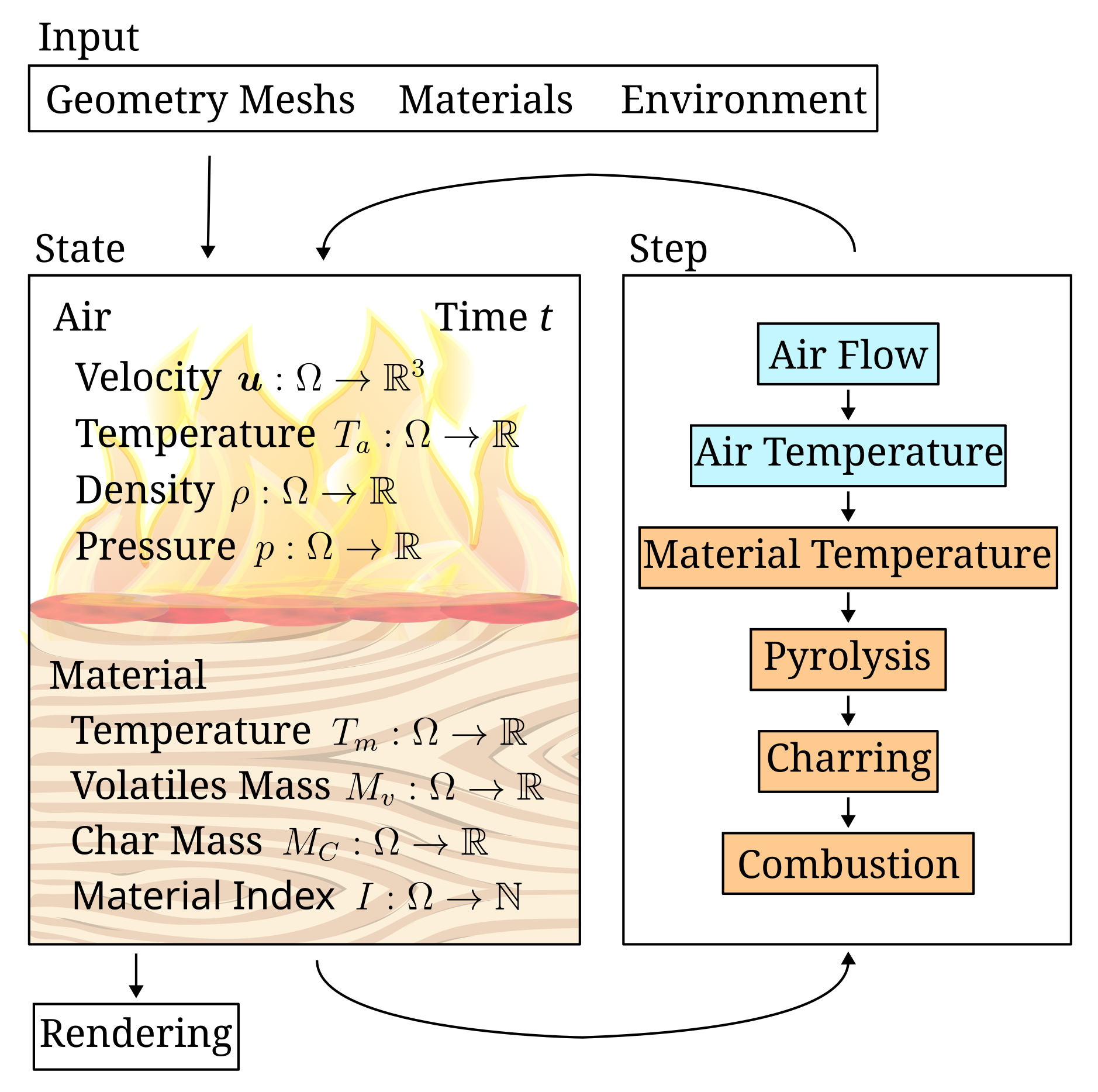}
\caption{\label{fig:overview}Overview of the presented \textsc{FlameForge} simulator.
The system's state is initialized from the input geometry, its materials and the environmental conditions, and consists of a single vector and seven scalar fields describing air and flames as well as the material.
In each update step, several sub-steps compute the next state of the system which is then forwarded to the rendering step to generate the visual output.}
\end{figure}

At the heart of our fire and combustion simulator -- illustrated in Figure~\ref{fig:overview} -- is its internal state representation, i.e., the overall configuration of the scene at a given point in time. The state is updated iteratively to progress the simulation through time.
The simulation takes place in a spatial domain $\Omega \subset \mathbb{R}^{3}$ that is filled with \emph{air} (that also includes other gases) and \emph{material} (burnable and non-burnable solids).

For each point in time $t\in\mathbb{R}$, the air is represented by fields for velocity $\boldsymbol{u}:\Omega\rightarrow\mathbb{R}^{3}$, temperature $T_{a}:\Omega\rightarrow\mathbb{R}$, density $\rho:\Omega\rightarrow\mathbb{R}$, and pressure $p:\Omega\rightarrow\mathbb{R}$.
Similarly, the material is represented by fields describing its temperature $T_{m}:\Omega\rightarrow\mathbb{R}$, the mass as a combination of volatiles $M_{v}:\Omega\rightarrow\mathbb{R}$ and char $M_{C}:\Omega\rightarrow\mathbb{R}$, as well as a material index $I:\Omega\rightarrow\mathbb{N}$ that is used as a lookup for heat transport and combustion properties.

The state for $t=0$ is initialized through the simulation input.
Each object in the scene is described through its surface by a triangle mesh with an assigned material index.
The input also contains additional environmental parameters such as ambient temperature or wind.
Internally, each field is represented by a multiresolution voxel grid.
Please note that for air and material in general different grid resolutions can be used.
The material mass is represented by a synchronized signed distance field (SDF) that allows for efficient distance queries during the char simulation.

During runtime, each step in the simulation maps the current state to the next one by evaluating the sub components:
The flow of air is computed through a fluid simulation.
Additionally, heat flow is computed where flames then naturally emerge as sufficiently hot areas of air.
We implement a bidirectional temperature coupling between air and material, such that flames can ignite material and combusting material produces flames.
Likewise, the heat flow through the material is computed and sufficiently hot parts undergo pyrolysis, creating a char layer and burnable volatiles.
After applying the insulation effect of the char, the material combusts, creating heat and smoke, which continue to be transported through the air flow.

\begin{figure*}
\includegraphics[width=0.245\textwidth]{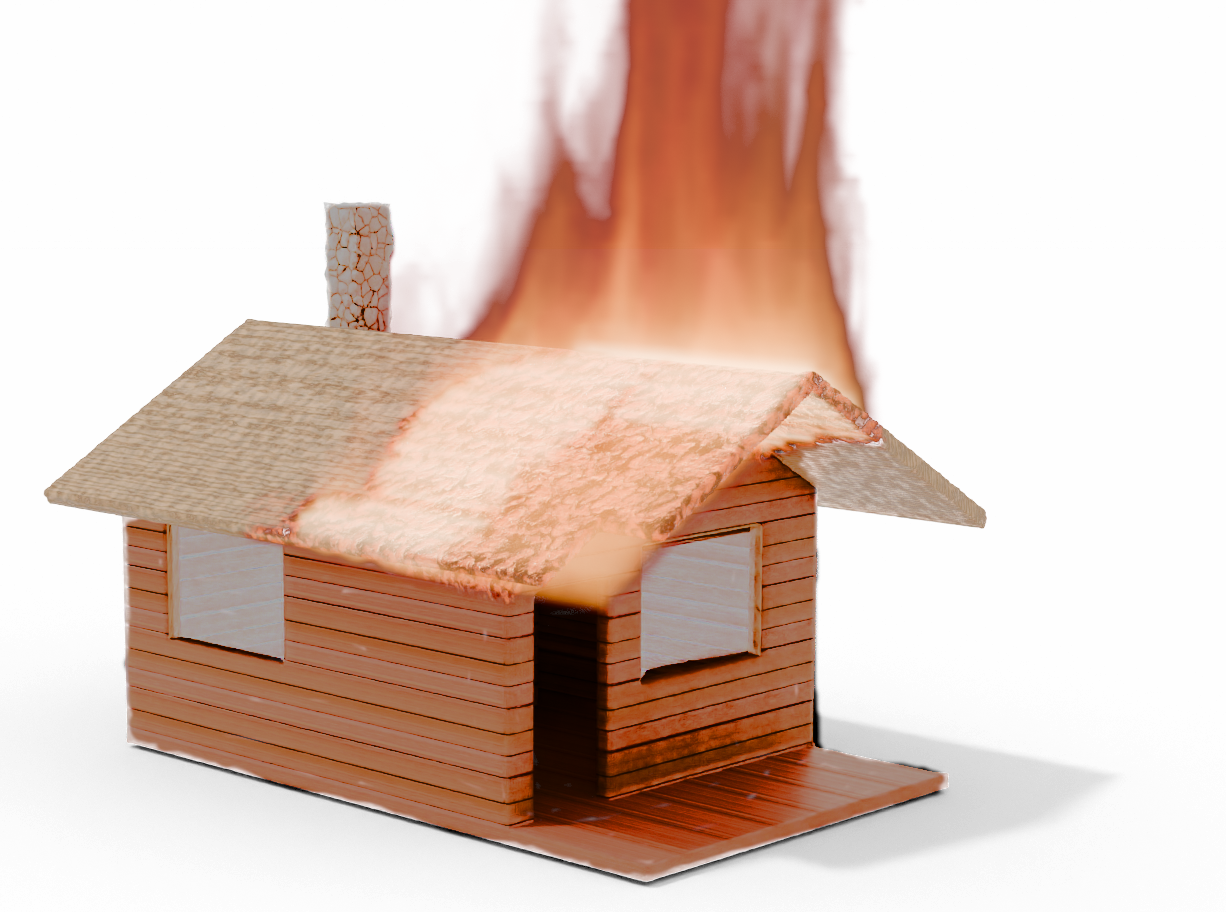}\hfill{}
\includegraphics[width=0.245\textwidth]{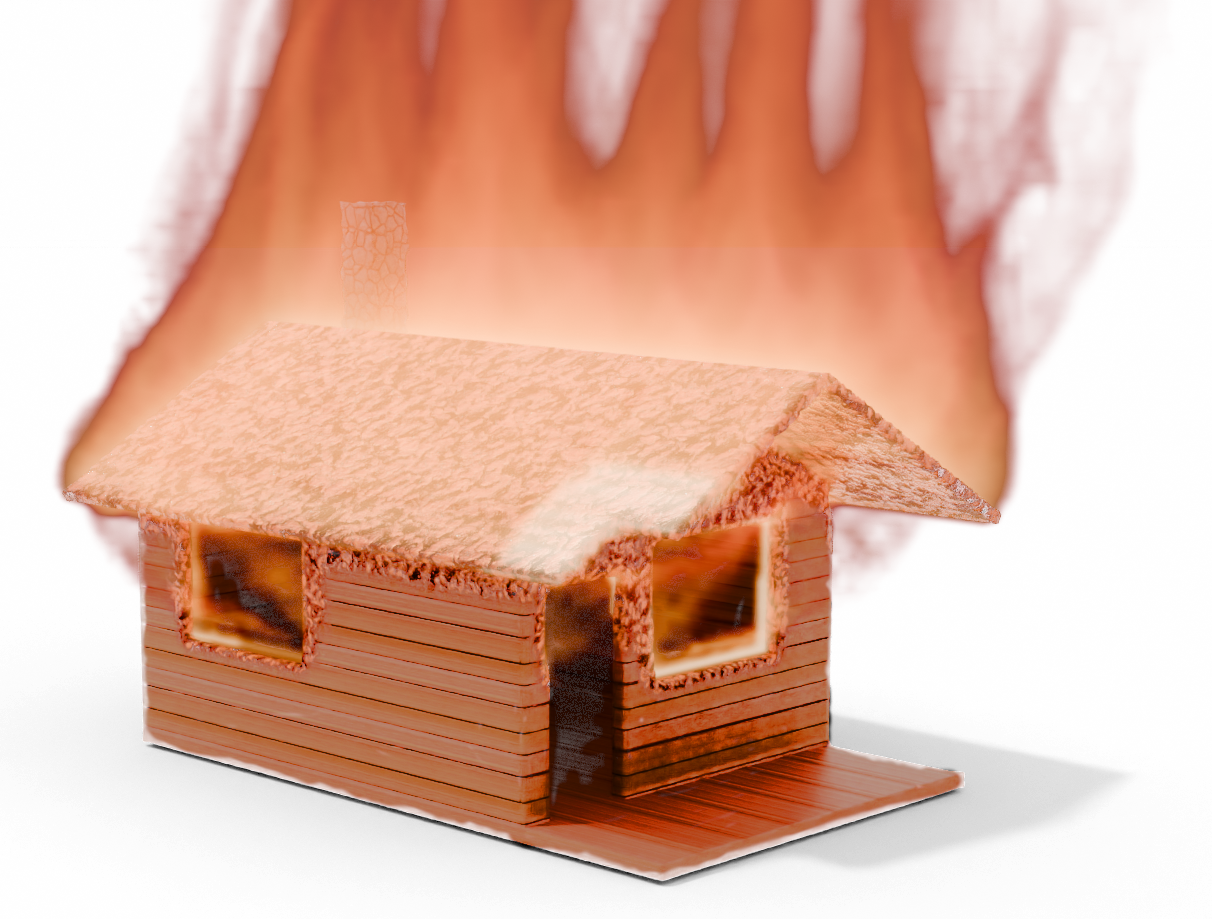}\hfill{}
\includegraphics[width=0.245\textwidth]{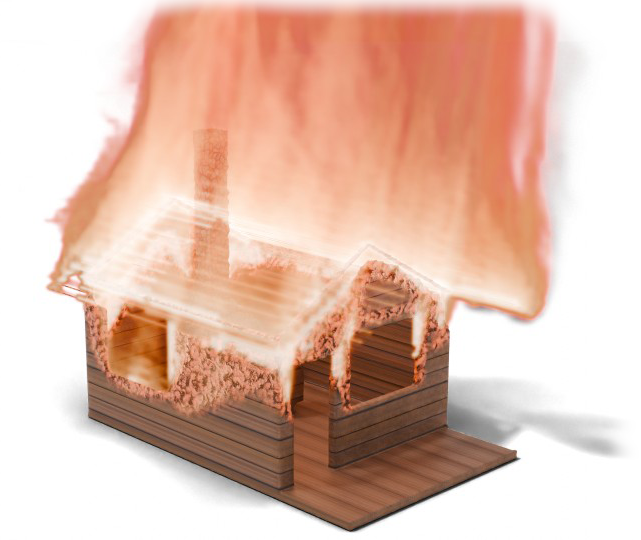}\hfill{}
\includegraphics[width=0.245\textwidth]{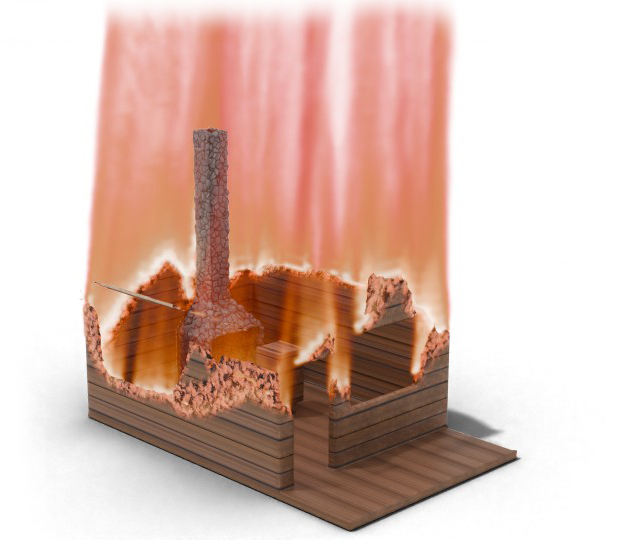} \\
\includegraphics[width=1.0\textwidth]{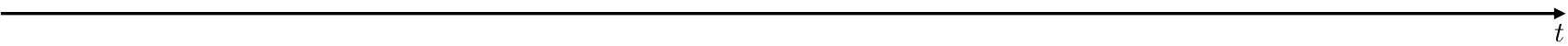}\\
\vspace{0.15cm}
\hfill{} $t_0$ \hfill{} \hfill{} $t_1$ \hfill{} \hfill{} $t_2$ \hfill{} \hfill{} $t_3$ \hfill{}
\vspace{-0.2cm}
\caption{\label{fig:result_materials}A scene featuring multiple materials simulating fire spread and combustion processes of a burning house which is composed of two types of wood (walls and roof are made up of different wood types), acrylic windows (non-charring), and a stone chimney (non-combustible). The roof catches fire ($t_0$) such as after a lightning strike causing smoke development and fire spread ($t_1$). After some time, the relatively thin wood of the roof has already been burned while the thick walls are barely touched ($t_2$). In contrast, the chimney made out of stone still remains intact at later stages ($t_3$) and even after the fire will be over.}
\end{figure*}




In the following, we describe our model through continuous differential equations.
Details about the numerical solvers for the resulting equations are presented in the upcoming Section~\ref{sec:Algorithmics}. An overview of all parameters and constants is provided in Section~\ref{sec:results}.

\subsection{Air and Fire Simulation}


\subsubsection{Air Flow}
\label{sec:airflow}
The fluid dynamics of the air are governed by the well known Navier Stokes equations.
Here, we use a simplified model for compressible flow, as introduced in \cite{nielsen2022physics}:
\begin{equation}
\frac{\partial\mathbf{u}}{\partial t}+\mathbf{u}\cdot\nabla\mathbf{u}=-\frac{\nabla p}{\rho}+\nu\nabla^{2}\mathbf{u}+\mathbf{b}\,,
\label{eq:NS}
\end{equation}
where the divergence is computed via 
\begin{equation}
\nabla\cdot\mathbf{u}=-\frac{1}{\rho}\frac{D\rho}{Dt}\,.
\label{eq:divergence}
\end{equation}
Here, $\nu$ is the kinematic viscosity of air.
The pressure $p$ is stored as the deviation from the hydrostatic pressure and is computed from the absolute pressure $\tilde{p}$ via $p=\tilde{p}-\rho_{\text{amb}}g$,
where $\rho_{\text{amb}}$ denotes the ambient air density and $g$ denotes earth's gravity (as a scalar quantity).
The buoyancy $\mathbf{b}=(0,0,b)^\mathsf{T}$ with $b=\left(1-\nicefrac{\rho_{\text{amb}}}{\rho}\right)g$ is applied as a force along the vertical axis.
The density $\rho$ is approximated via $\rho\approx\rho_{\text{amb}}\,{T_{\text{amb}}}/{T_{a}}$ (where $T_{\text{amb}}$ is the ambient air temperature \cite{nielsen2022physics}).

The air flow uses the material geometry as boundary conditions (see Section~\ref{sec:air_impl}).

\subsubsection{Temperature}
The air temperature is modeled according to Tritton et al.~\shortcite{tritton2012physical}:
\begin{equation}
\frac{\partial T_{a}}{\partial t}+\mathbf{u}\cdot\nabla T_{a}=k\nabla^{2}T_{a}+\gamma_{a}(T_\text{amb}^{4}-T_{a}^{4})+S_{T_a}\,,
\label{eq:atmosphere_heat_transfer}
\end{equation}
where $k$ is the thermal diffusivity of air, $\gamma_a$ is the radiative cooling coefficient of air, and $S_{T_a}$ is a source term for temperature.
We compute $\gamma_a$ according to the Stefan-Boltzmann law as
\[
\gamma_{a}=\frac{\varepsilon\sigma A}{C_{a} M_{a}}\,,
\]
where $\varepsilon$ is the black body emission rate, $\sigma$ is the Boltzmann constant, $A$ is the surface area of a voxel, $C_a$ is the heat capacity of air, and $M_a$ is the mass of a voxel cell.
Since the light emission of flames mostly stems from black body radiation, we approximate $\varepsilon=1$ (ideal black body).

\subsubsection{Smoke}
While flames are modeled as hot air, the density of smoke at each position is independent from the air.
In our model, we treat smoke mostly as a visual effect.
It is created during combustion and moves together with its surrounding air but does not affect any other part of the simulation itself.
The smoke field $S$ is described via 
\begin{equation}
\frac{\partial S}{\partial t}+\mathbf{u}\cdot\nabla S=S_{S_a}\,,
\label{eq:smoke_advec}
\end{equation}
where $S_{S_a}$ is a source term for smoke.

\subsection{Material Combustion}


\subsubsection{Material Temperature}
The change of the material temperature  $T_m$ is a combination of thermal diffusion, radiative cooling and heat sources:
\begin{equation}
\frac{\partial T_{m}}{\partial t}=\beta\nabla^{2}T_{m}+\gamma_{m}(T_\text{amb}^{4}-T_{m}^{4})+S_{T_{m}}\,.
\label{eq:solid_combustion}
\end{equation}
Here, $\beta$ is the thermal diffusivity, $\gamma_m$ is the radiative cooling coefficient for material (computed as above), and $S_{T_m}$ is a temperature source term.
Heat is also exchanged between air and material:
\begin{equation}
\label{eq:heat_exchange}
\frac{\partial T_{m}}{\partial t}=\left(T_{a}-T_{m}\right)\phi_{m}\,,
\,\,\,\,\,\,\,
\frac{\partial T_{a}}{\partial t}=\left(T_{m}-T_{a}\right)\phi_{a}\,,
\end{equation}
where $\phi_m$ and $\phi_a$ are the exchange rates between air and material and vice versa.
They are computed from the general exchange rate $\phi$ via
\begin{equation}
\phi_{m}=\frac{\phi}{C_{m}M_{m}} \hspace{1em} \text{and} \hspace{1em} \phi_{a}=\frac{\phi}{C_{a}M_{a}}\,,
\end{equation}
where $C_m, C_a$ are the heat capacities of the material and air, and $M_m, C_m$ are the masses of an air and material voxel.


\subsubsection{Pyrolysis}
The process of thermal decomposition of the material is denoted as pyrolysis.
In general, this is a highly complex process and the products depend on the pyrolysed material and the environmental conditions.
Here, we use a simplified model, the so called \emph{one-step global model} \cite{di1993modeling}.
It assumes that a voxel of sufficiently hot material is immediately decomposed into volatiles (burnable gases) and char:
\[
\text{Solid Fuel} \xrightarrow{\text{Heat}} \text{Volatiles} + \text{Char}\,.
\]
The char will also eventually burn, but has a far lower combustion rate than the volatiles.
Non-charring materials will not produce chars during pyrolysis, but only volatiles.
Since the model assumes, that the pyrolysis process happens instantaneously once the ignition threshold $T_{m_0}$ is passed, our representation does not need to distinguish the before and after state.
Therefore, we directly store the char mass $M_c$ and the volatile mass $M_v$ for each grid cell.
We also store relative mass, rather than absolute mass.
This means, that all fields can be initialized with $1$.
Combustion parameters like $\epsilon_v$ (the amount of heat produced per unit mass volatiles burned) reflect this accordingly.
This design reduces the complexity without limiting the effects that can be simulated.

\subsubsection{Charring}
The pyrolysis generates char, which eventually forms a char layer with increasing thickness.
Below this layer, oxygen becomes less available and the combustion rate slows down.
We use the char insulation model as described by Pirk et al.~\shortcite{pirk2017interactive} to compute the char insulation coefficient $c \in [0, 1]$ via 
\begin{equation}
c = c_{\mathsf{min}}+\left(1-c_{\mathsf{min}}\right) \mathsf{exp}(-h c_{r})\,.
\end{equation}
Here, $h$ is the thickness of the char layer, $c_r$ the char insulation rate and $c_\text{min}$ is the minimum insulation value. 

The accurate computation of the char insulation for volumetric structures would require solving an integral over all possible paths from an inner voxel to the materials surface and computing the insulation effect along all these paths.
We simplify this problem by computing an SDF from the voxel volume and use it to query the distance $h$ of a voxel from the material's boundary.
Since the geometric shape of the material changes during the combustion process, this SDF is iteratively updated during the simulation.

Please note that this insulation effect is not an effect unique to the material properties of char.
Therefore, virgin wood (before the pyrolysis) and voxels containing only char are treated in the same way.
Increasing the realism of the insulation effect would start with employing a more complex pyrolysis model \cite{di1993modeling}.

\subsubsection{Combustion}
Burning volatiles and char lose mass according to their individual combustion rates $\epsilon$, the char insulation $c$ and a temperature dependent reaction rate $\xi$:
\begin{align}
\label{eq:mass_loss}
\frac{\partial\text{\ensuremath{M_{c}}}}{\partial t}=\epsilon_{c} c \xi \left(T_m \right)\,,
\,\,\,\,\,\,\,
\frac{\partial\text{\ensuremath{M_{v}}}}{\partial t}=\epsilon_{v} c \xi \left(T_m \right)\,,\\
\text{where}\,\,\,\xi\left(T_{m}\right)=f\left(\frac{T_{m}-T_{m_{0}}}{T_{m_{1}}-T_{m_{0}}}\right)\,\,\,\text{for}\,\,\,T_{m_{0}}\leq T_{m}\leq T_{m_{1}},
\end{align}
and $\xi\left(T_{m}\right)=0$ for $T_{m}<T_{m_{0}}$ and $\xi\left(T_{m}\right)=1$ for $T_{m}>T_{m_{1}}$. The function $f\left(x\right)=3x^{2}-2x^{3}$ describes a smooth interpolation, and $T_{m_{0}}$ and $T_{m_{1}}$ denote minimum and maximum temperature limits.


Simultaneously, heat ($S_{T_m}$) and smoke ($S_{S_m}$) are generated by the reaction, in linear dependence on the amount of mass consumed:
\begin{equation}
S_{T_m}=-T_{M_{c}}\frac{\partial M_{c}}{\partial t}-T_{M_{v}}\frac{\partial M_{v}}{\partial t}\,,
\,\,\,
S_{S_m}=-S_{M_{c}}\frac{\partial M_{c}}{\partial t}-S_{M_{v}}\frac{\partial M_{v}}{\partial t}\,.
\label{eq:tm_source}
\end{equation}
Since $M_{v}$ and $M_{c}$ store relative quantities, $T_{M}$ describes the energy released per volume and not per mass.
As smoke is created in a material voxel but travels through air, the material smoke source $S_{S_m}$ is then projected to air smoke sources $S_{S_a}$ in neighboring voxels.


\section{Algorithmics}
\label{sec:Algorithmics}

\begin{figure}
\includegraphics[width=0.95\columnwidth]{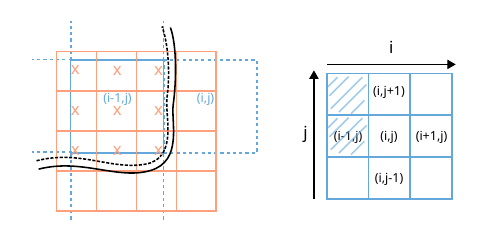}
\vspace{-0.5cm}
\caption{\label{fig:grid_projection}
Illustration of the grid projection in 2D.
Left: Fine (orange) to coarse (blue). Black Lines: Material contour. Orange x: Voxels inside the material.
Right: Neighborhood look up. Blue dashes: Material occupied voxels.}
\end{figure}

The mathematical model described in the previous section provides the basis for the implementation of
our \textsc{FlameForge} simulator. In the following, we provide details for the numerical integration procedure including the setup of multiresolution and adaptive data structures and necessary details about the numerical integration procedure. The overall approach is summarized in Algorithm \ref{alg:pseudo_code} which has been implemented in C++ using OpenVDB \cite{museth2013vdb}.

\subsection{Multiresolution and Adaptive Data Structures}

To increase the efficiency of computation, we employ an adaptive multi-grid approach.
Many geometric structures are sparse, such as buildings where the walls are thin compared to their interior.
By using adaptive voxel grids, storage and computation time are saved by efficiently marking large areas of the simulation domain as empty.
Additionally, the air flow simulation uses a much lower (usually $5\times$) resolution than the material representation.
This is motivated by the fact that flame movement is generally more volatile and diffuse, and temperature is transported through convection and conduction more easily than through matter.
Therefore, less spatial detail is required.

We use the multiresolution voxel grids implementation of OpenVDB (\texttt{openvdb::FloatGrid}), which transparently handles masking and interpolation.
The SDF relies on the same grid implementation as the underlying data structure.
Its values are computed through the fast sweeping method \cite{zhao2005fast}.

\subsection{Numerical Integration}
\label{sec:air_impl}

In the following, we disclose the relevant numerical details of our \textsc{FlameForge} simulator.

\begin{figure}
\centering
\includegraphics[width=0.495\columnwidth]{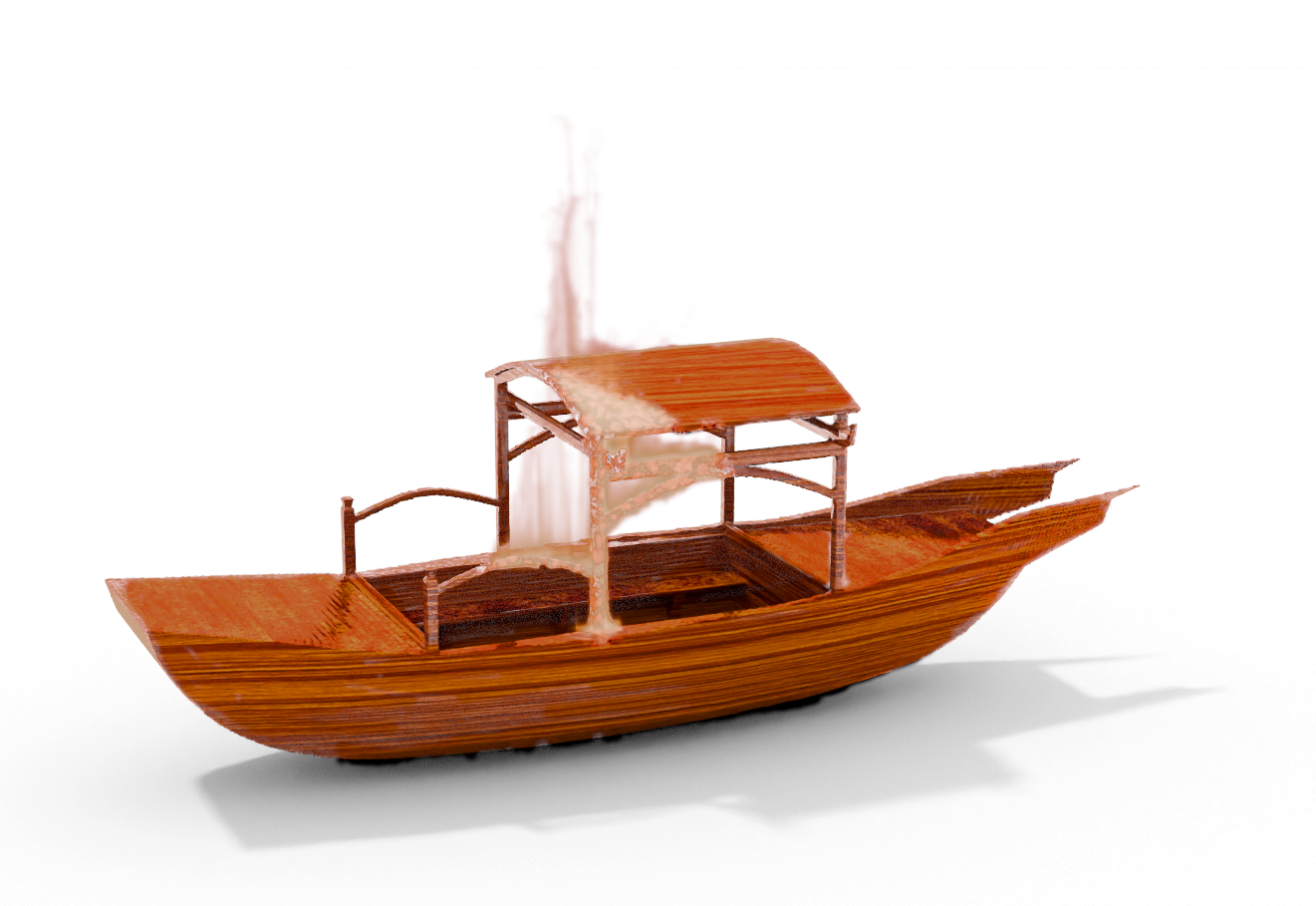}
\includegraphics[width=0.495\columnwidth]{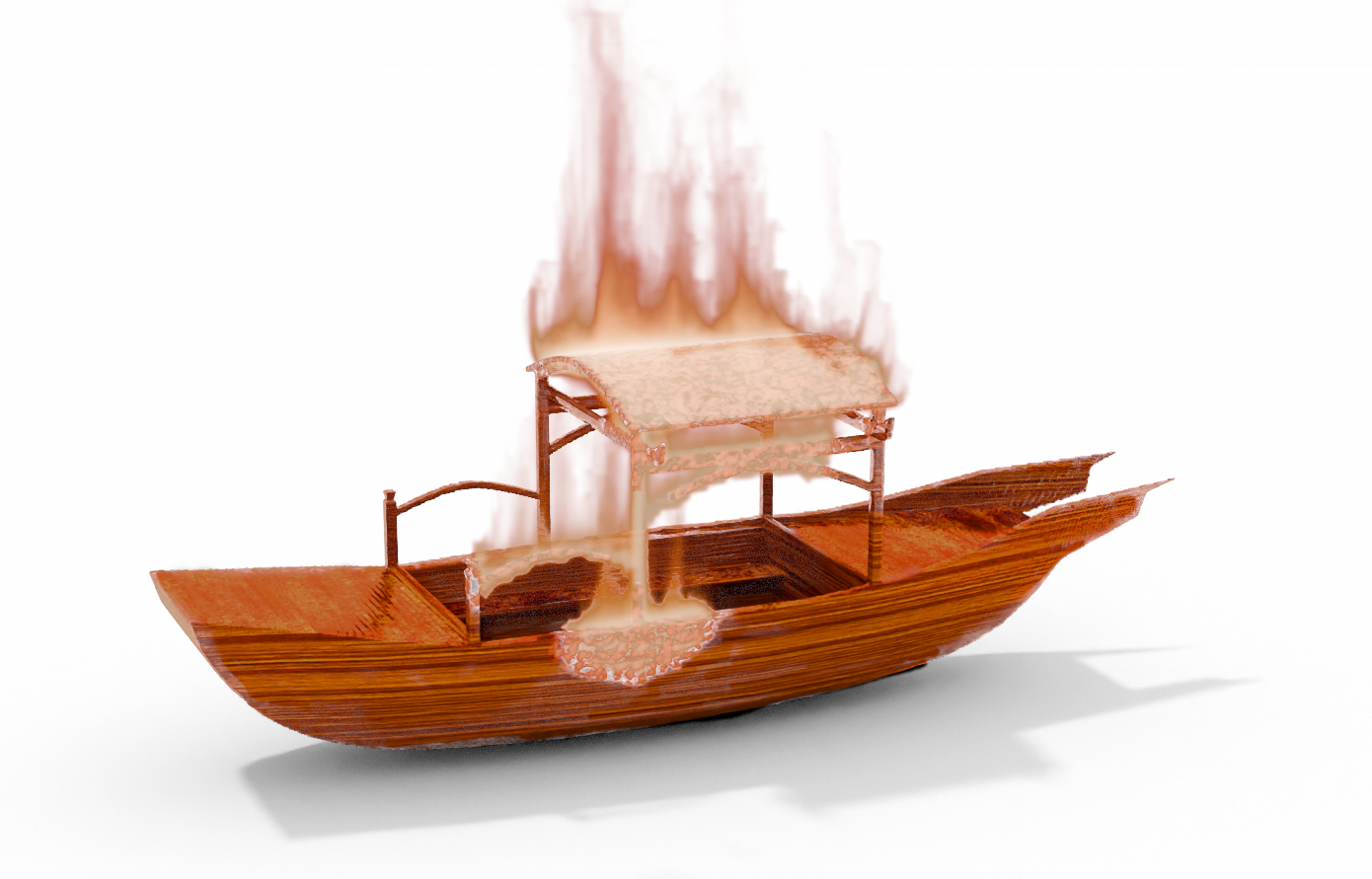}
\includegraphics[width=0.495\columnwidth]{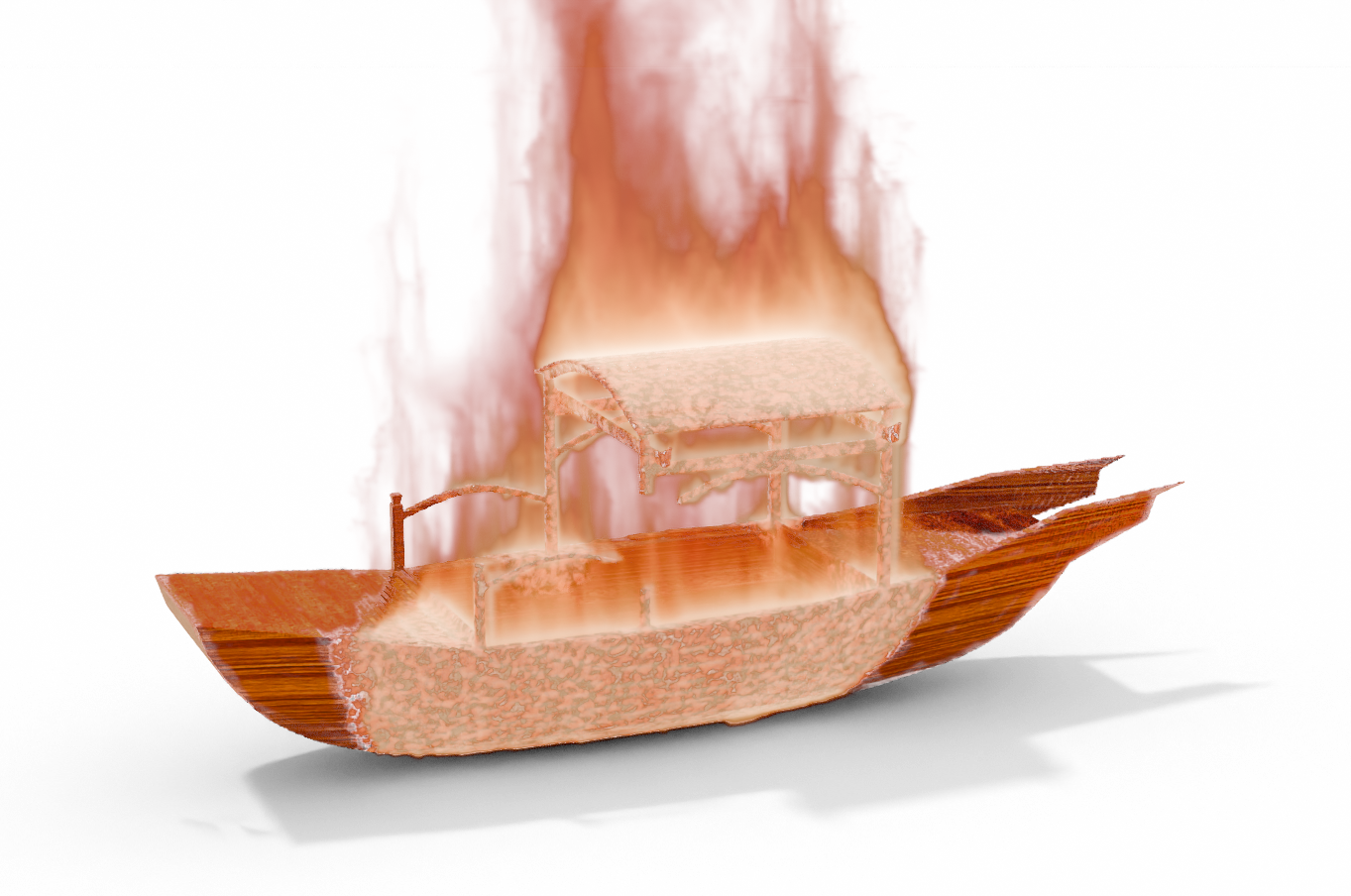}
\includegraphics[width=0.495\columnwidth]{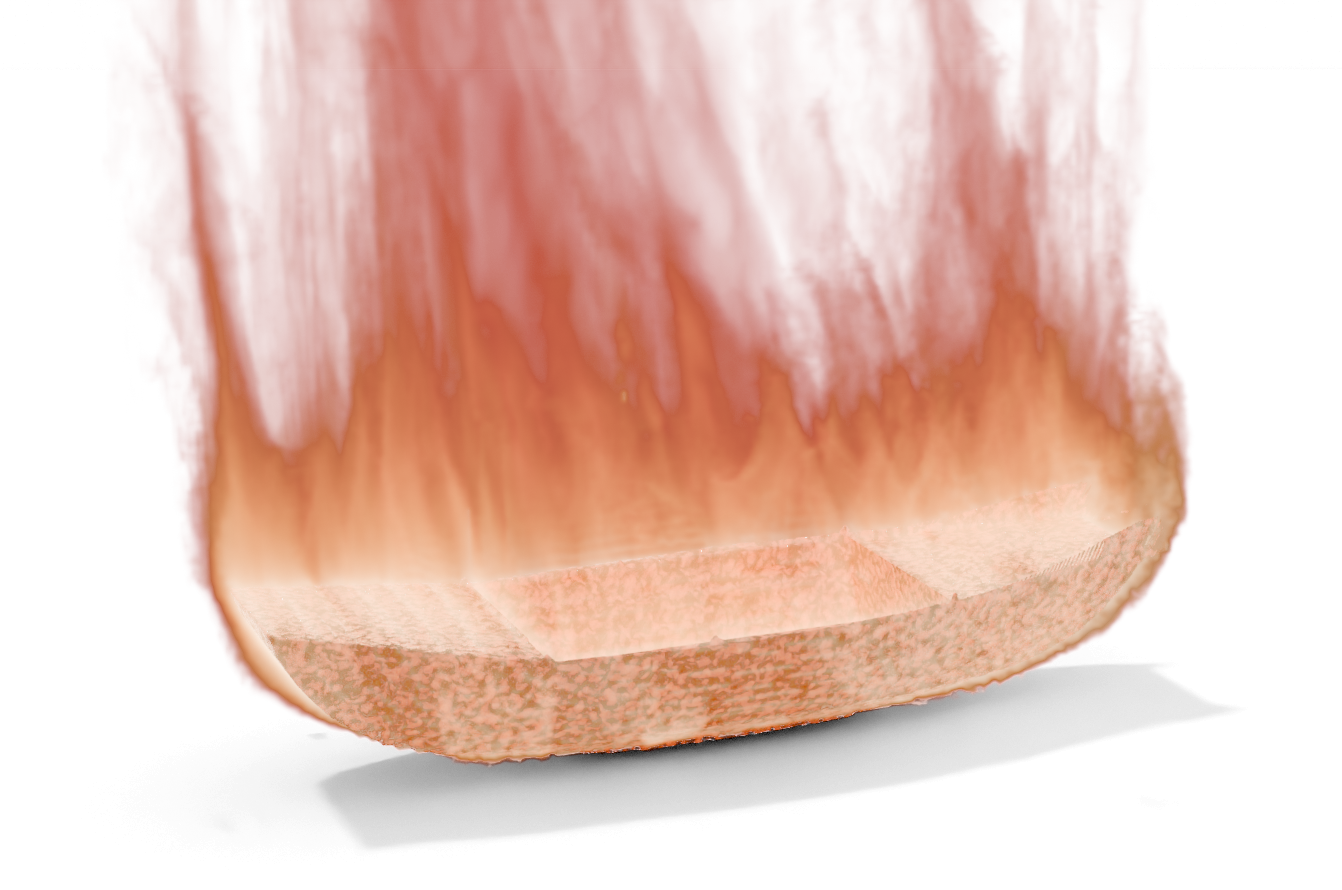} 
\vspace{-0.4cm}
\caption{\label{fig:result_basic}Illustration of fire spread and combustion using complex geometry: A traditional, relatively flat-bottomed Chinese boat fully made out of wood has been ignited at the bottom (upper left). The fire then spreads quickly to the roof (upper right) and further progresses (lower left) before the whole boat burns brightly (lower right).}
\end{figure}

\subsubsection{Fluid Solver}
The Eulerian grid-based method is used to solve the equations above, which are discretized in a staggered way on the grid \cite{bridson2015fluid}.
Scalar variables (such as pressure,  temperature, mass, etc.) are located at the centers of voxels, and velocity components  are located at the face centers of the voxels.
The time derivatives are discretized with a forward difference scheme and the Laplacian diffusion terms are discretized with a finite central difference scheme.
The advection term is solved using the MacCormack scheme \cite{selle2008unconditionally} and the Poisson equations in the projection step are solved with AMGCL \cite{demidov2019amgcl}.
In addition, we explicitly applied slip solid boundaries to handle solid materials \cite{bridson2015fluid}.

\subsubsection{Grid Projection}
\label{sec:grid_pro}
In Eq.~(\ref{eq:heat_exchange}), the temperature difference between material ($T_m$) and air ($T_a$) is computed for material boundary voxels when transferring heat from material to air.
However, both quantities are defined on different voxel grids using different resolutions.
Furthermore, the material grid stores values only on the inside of the geometry, while the air grid stores values only on its outside, see Figure~\ref{fig:grid_projection}.
We therefore project the material temperature $T_m$ onto the empty cells of the air temperature grid $T_a$ on the inside boundary of the geometry.
For each inside boundary cell of $T_a$ (e.g., $(i-1, j)$), the overlapping voxels of $T_m$ are determined and their maximum value is used as the value of $T_a$.
Then, each outside boundary voxel (e.g., $(i, j)$) accumulates the values of the inside boundary voxels in its neighborhood by summing them up.
We now have $T_m$ and $T_a$ defined at the same voxel cell and can compute Eq.~(\ref{eq:heat_exchange}).

To exchange heat from air to material, we follow the reverse procedure.
The inside boundary voxels of $T_a$ accumulate the values from their neighbors.
Then, trilinear interpolation is used to project these values to the high resolution grid of $T_m$.




\begin{algorithm}
\SetAlgoLined
\LinesNumbered
\KwIn{Current system state.}
\KwOut{Updated system state.}

\textbf{for each} air grid cell \textbf{do} \\
\,\,|\,\,\, Update velocity by pressure projection, advection, and viscosity diffusion, according to Eq.~(\ref{eq:NS}) and Eq.~(\ref{eq:divergence}). \\
\,\,|\,\,\, Update velocity by external gravity and buoyancy force, according to Section~\ref{sec:airflow}.\\
\,\,|\,\,\, Update temperature by advection, diffusion, and radiative cooling according to Eq.~(\ref{eq:atmosphere_heat_transfer}).\\
\,\,|\,\,\, Update smoke by advection according to Eq. (\ref{eq:smoke_advec}).\\
\textbf{end}\\

\textbf{for each} interface grid cell between material and air \textbf{do} \\
\,\,|\,\,\, Temperature exchange according to Eq.~(\ref{eq:heat_exchange}).\\
\textbf{end}

\textbf{for each} overlapped air grid cell \textbf{do} \\
\,\,|\,\,\, Update the temperature and smoke of the air grid cell with the overlapped solid material cells as described in Section~\ref{sec:grid_pro}.\\
\textbf{end}\\

\textbf{for each} material grid cell \textbf{do} \\
\,\,|\,\,\, Update temperature by diffusion and radiative cooling according to Eq.~(\ref{eq:solid_combustion}) . \\
\,\,|\,\,\, Update volatile and char mass, and calculate mass loss rate according to Eq.~(\ref{eq:mass_loss}).\\
\,\,|\,\,\, Calculate the heat and smoke source term according to Eq.~(\ref{eq:tm_source}). \\
\textbf{end}\\

Update SDF based on the material mass field.
\\
Update the grid mask of the simulation domain. \\

\caption{\textsc{FlameForge}'s time integration procedure.}
\label{alg:pseudo_code}
\end{algorithm}
\section{Numerical Examples}
\label{sec:results}

We showcase the capabilities of our \textsc{FlameForge} framework through a variety of simulations. Table~\ref{tab:parameters} provides an overview of the relevant parameters including numerical values and physical units. All numerical experiments have been carried out on a workstation with an Intel\textsuperscript{\textregistered} Xeon\textsuperscript{\textregistered} CPU E5-2699 and 256~GB memory. A performance statistics is provided in Table~\ref{table:performance}. The final results are visualized using the Cycles renderer within Blender~4.0.

\begin{table}
\centering
\caption{\label{tab:parameters}Summary of relevant parameters used within the presented model. The parameter values of the quantities listed in the first group (i.e., listed above the first dashed horizontal line) are well known natural constants~\cite{birch1942handbook}. The parameters listed in the second group (i.e., listed between the two dashed horizontal lines) are adapted according to Pirk et al.~\shortcite{pirk2017interactive}. The remaining two parameters can be found in the work of MacLeod et al.~\shortcite{macleod2023quantifying}.}
\scriptsize{
\begin{tabular}{llll}
\hline
\textbf{Identifier} & \textbf{Description} & \textbf{Value (Range)} & \textbf{Unit}\tabularnewline
\hline 
$\nu$ & Kinematic viscosity of air & $1.6\cdot10^{-5}$ & $\text{m}^2\text{s}^{-1}$ \tabularnewline
$g$ & Gravity (earth) & 9.81 & $\text{m}\text{s}^{-2}$ \tabularnewline
$\rho_{a}$ & Density (air) &  1.2041 & $\text{kg}\,\text{m}^{-3}$ \tabularnewline
$T_{\text{amb}}$ & Ambient temperature &  293 & $\text{K}$ \tabularnewline
$k$ & Thermal diffusivity coefficient (air) &  $1.8\cdot10^{-5}$ & $\text{m}^2\text{s}^{-1}$ \tabularnewline
$\gamma_{a}$ & Radiative cooling coefficient (air) & $5.0\cdot10^{-11}$ & $\text{K}^{-3} \text{s}^{-1}$ \tabularnewline
$\beta$ & Thermal diffusivity &  $0.82\cdot10^{-7}$ & $\text{m}^2\text{s}^{-1}$\tabularnewline
\hdashline
$\gamma_{m}$ & Radiative cooling coefficient & $5.9\cdot10^{-14}$ & $\text{K}^{-3} \text{s}^{-1}$\tabularnewline
$\phi_{m}$ & Thermal exchange rate (air-material) & $2.0 \cdot 10^{-2}$ & $\text{s}^{-1}$\tabularnewline
$\phi_{a}$ & Thermal exchange rate (material-air) & $5.0 \cdot 10^{-2}$ & $\text{s}^{-1}$\tabularnewline
$T_{m_{0}}$ & Lower threshold (pyrolysis and combustion) & $[150,280]$ & \text{K} \tabularnewline
$T_{m_{1}}$ & Temperature limit (maximum combustion rate) & $[400, 500]$ & \text{K} \tabularnewline
$C_{\mathsf{min}}$ & Maximum value (insulation) & 0.1 & 1 \tabularnewline
$C_{r}$ & Rate (char insulation) &  75.0 & $\text{m}^{-1}$ \tabularnewline
$\epsilon_{c}$ & Mass loss rate (char combustion) & $[0.1,1.0]\cdot10^{-3}$ & $\text{kg}\,\text{s}^{-1}$ \tabularnewline
$S_{M_{c}}$ & Smoke generation rate (char) &  $1.0\cdot10^{3}$ & $\text{kg}^{-1}$\tabularnewline
$S_{M_{v}}$ & Smoke generation rate (volatiles) &  $1.0\cdot10^{3}$ & $\text{kg}^{-1}$\tabularnewline
$\epsilon_{v}$ & Mass loss rate (volatiles combustion) &  0.1 & $\text{kg}\,\text{s}^{-1}$ \tabularnewline
\hdashline
$T_{M_{c}}$ & Heat generation rate (char) &  $3.0\cdot10^{7}$ & $\text{kg}^{-1} \text{K}$ \tabularnewline
$T_{M_{v}}$ & Heat generation rate (volatiles) & $2.0\cdot10^{7}$ & $\text{kg}^{-1} \text{K}$ \tabularnewline
\hline
\end{tabular}}
\end{table}

\begin{table}
\centering
\caption{Performance statistics and scene parameters for all simulation results. Please note that the grid resolution denotes the one used for the fluid (both air flow and fire) simulation. The combustion simulation is performed on the grid that is used to implicitly represent the solid objects requiring additional computational resources. The computation time includes both fluid and combustion simulations and is provided per frame.}
\label{table:performance}
\scriptsize{
\begin{tabular}{l l l l l}
\hline
\textbf{Figure} & \textbf{Scene} & \textbf{Grid Resolution}  & \textbf{Runtime} & \textbf{\#Frames}\\
\hline
\ref{fig:result_materials} & House (multiple materials) &$110 \times 100 \times 70$& 6.7s& 8\,818\\
\hdashline
\ref{fig:result_basic} & Chinese Boat & $260 \times 120 \times 120$ & 30s & 2\,961 \\
\hdashline
\ref{fig:result_tree_house} (top)& Tree and House (wind) & $120 \times 80 \times 70$ &  4.7s & 4\,323 \\
\ref{fig:result_tree_house} (bottom) & Tree and House (no wind) & $120 \times 80 \times 70$ &   4.7s& 2\,339\\
\hdashline
\ref{fig:experiment} & Cube (non-charring) & $30 \times 30 \times 30$& 1.1s &37\\
\ref{fig:experiment} & Cube (charring) & $30 \times 30 \times 30$&  1.1s &976\\
\hdashline
\ref{fig:result_tower} (top) & Tower (roof catches fire) &  $70 \times 150 \times 70$& 4.7s & 24\,350\\
\ref{fig:result_tower} (middle) & Tower (flamethrower) & $100 \times 150 \times 70$& 5.8s & 4\,557\\
\ref{fig:result_tower} (bottom) & Tower (fire source below) & $70 \times 150 \times 70$ & 4.5s &15\,070\\
\hline
\end{tabular}}
\end{table}

\subsection{Matter Combustion}

We demonstrate the capability of our system to handle complex geometry by igniting a model of a traditional Chinese boat, made entirely out of wood, see Figure~\ref{fig:result_basic}.
The model consists of bulky elements such as the front, and more intricate details, such as the beam construction under the roof, and thus benefits from the adaptive grid.
Furthermore, the roof acts as a complex obstacle for the fluid simulation of the air.

\subsection{Different Materials}
The ability to handle different materials is shown in Figure~\ref{fig:result_materials}.
The house consists of a total of four different materials:
Hard, slow burning wood for the walls and floor, lighter, but faster burning wood for the roof, non-charring acrylic glass for the windows and a chimney made out of stone which is entirely unaffected by combustion but still blocks air and transmits heat.

As the simulation progresses, the fire spreads through different parts of the house at different rates, according to the involved materials.
In the end, only the stony chimney and the thick ground plate of the house remain (the fire extinguishes when the specific conditions do not allow to keep the local temperature above the ignition point).

\subsection{Charring / Experimental Verification}

\begin{figure*}
\includegraphics[width=0.235\textwidth] {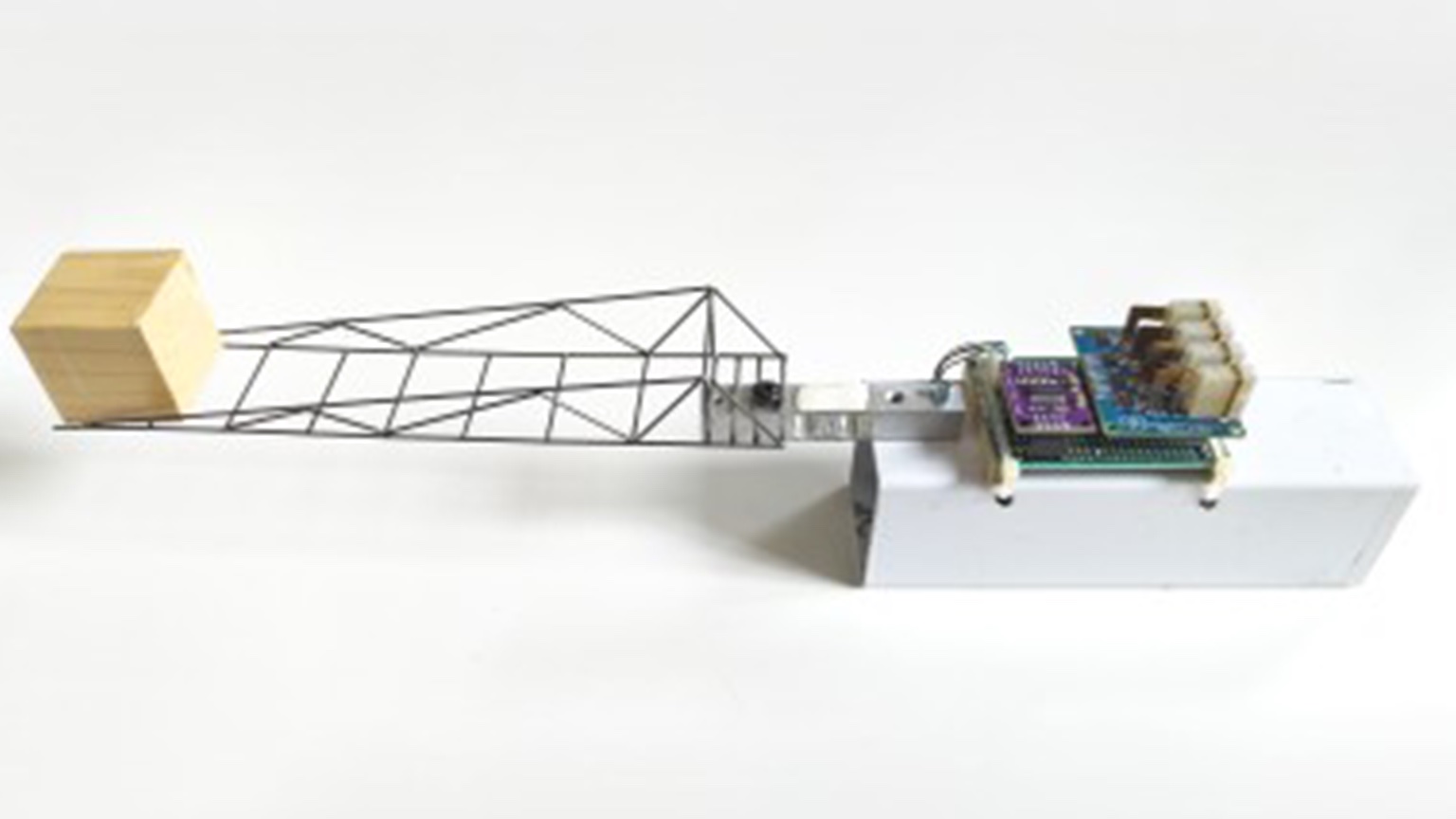}\hfill{}
\includegraphics[width=0.235\textwidth] {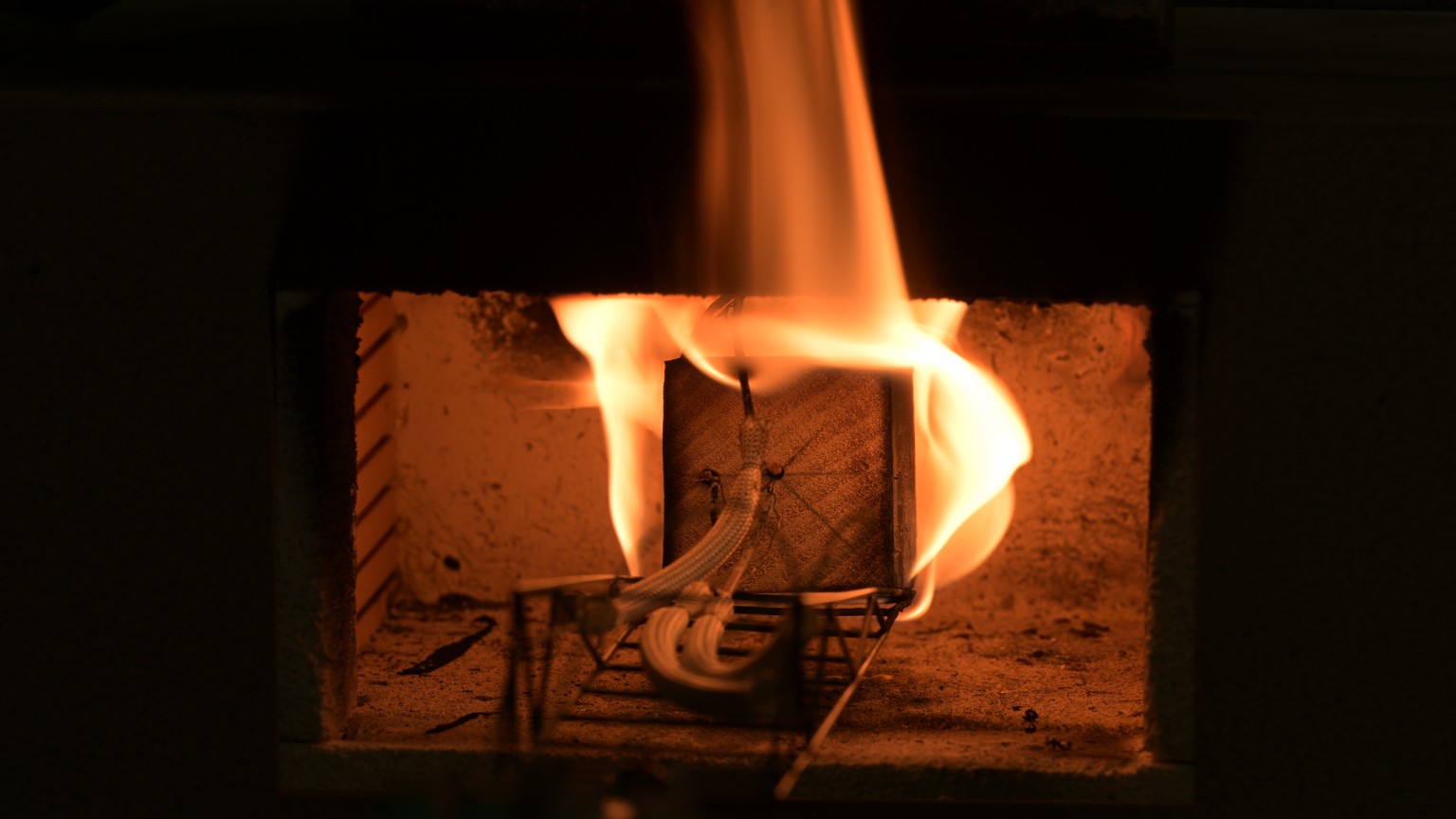}\hfill{}
\includegraphics[width=0.235\textwidth] {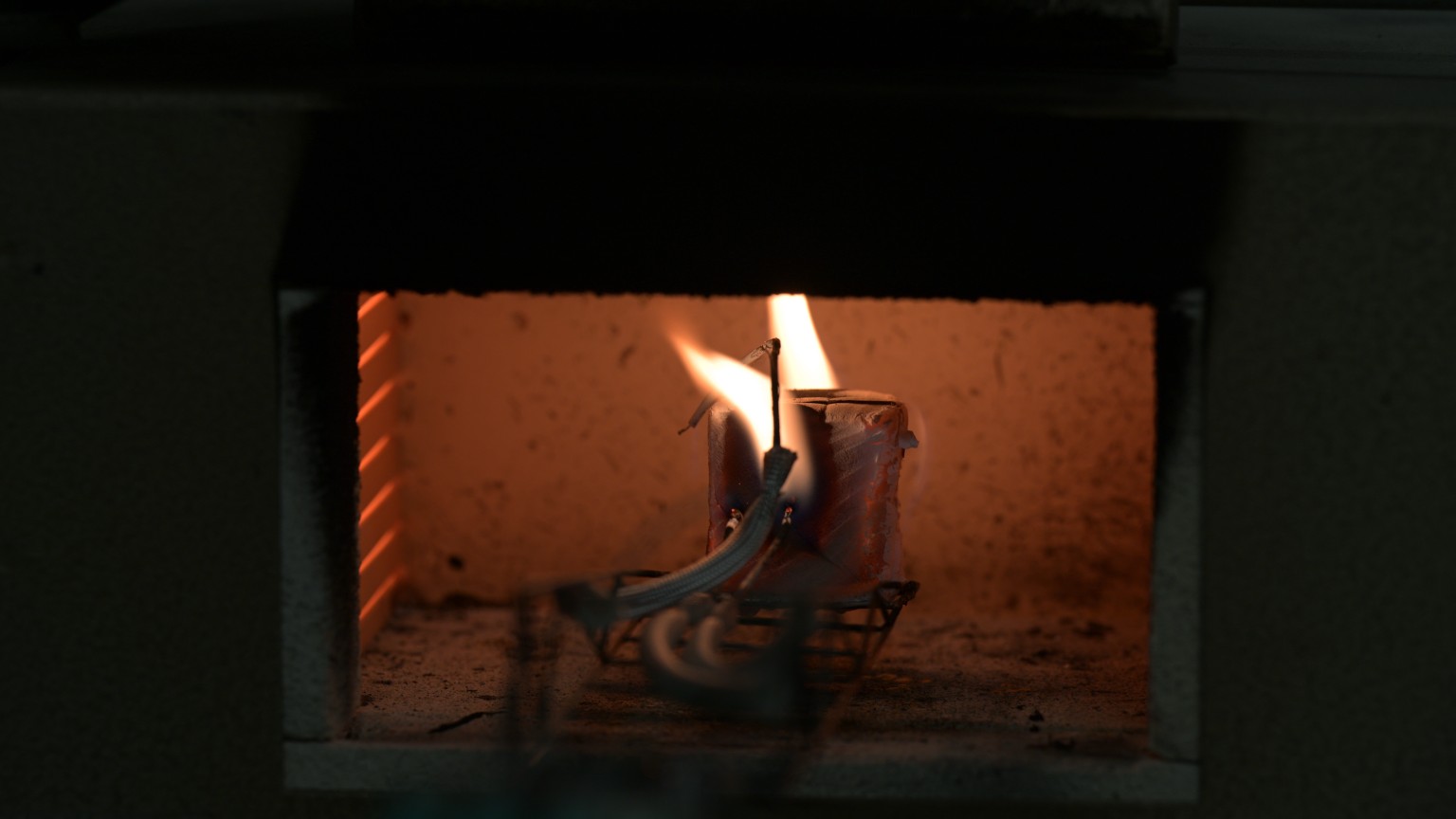}\hfill{}
\includegraphics[width=0.235\textwidth] {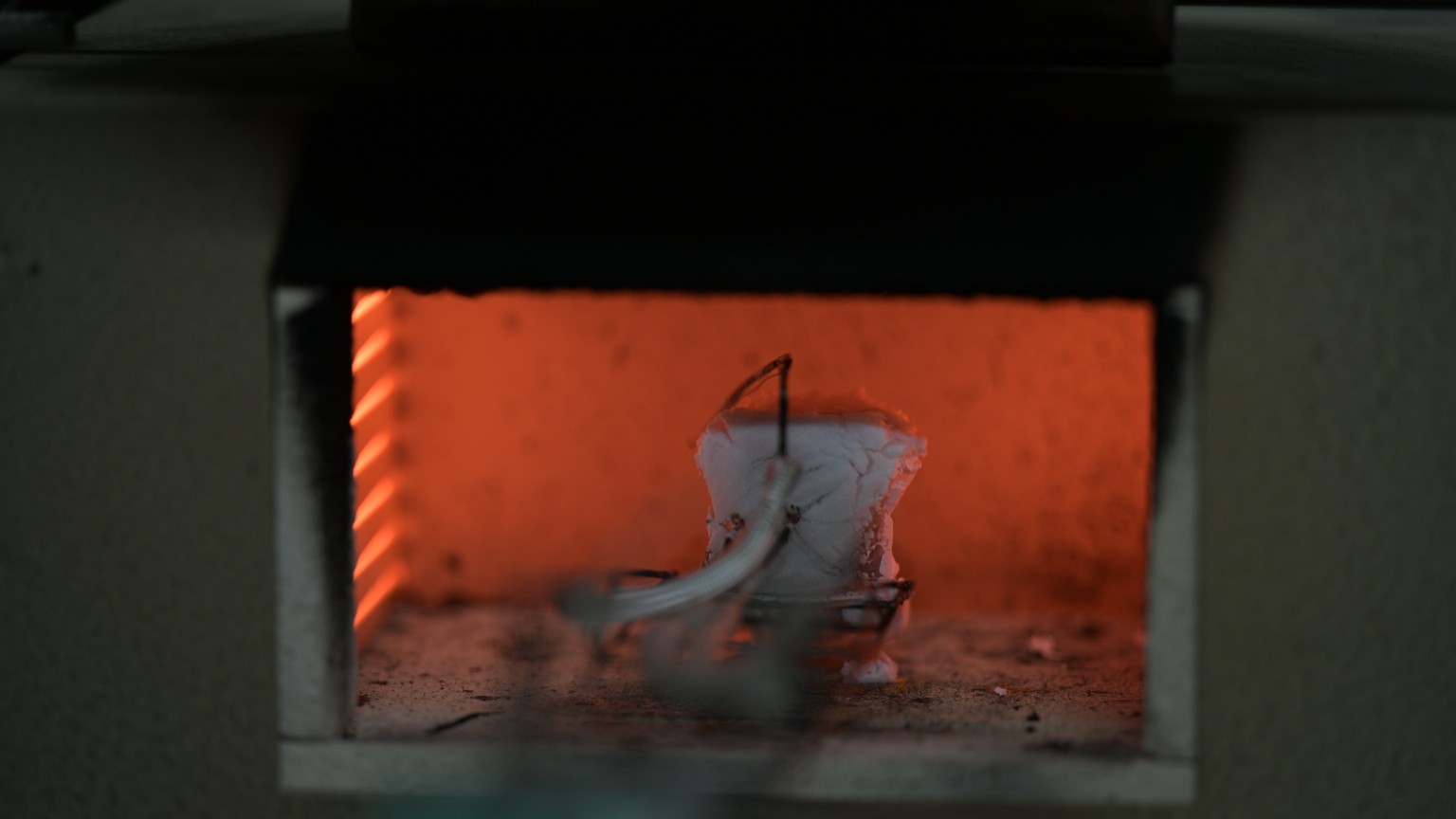} \hfill{} \rotatebox[x=0em, y=5em]{-90}{Wood}\\
\includegraphics[width=0.235\textwidth] {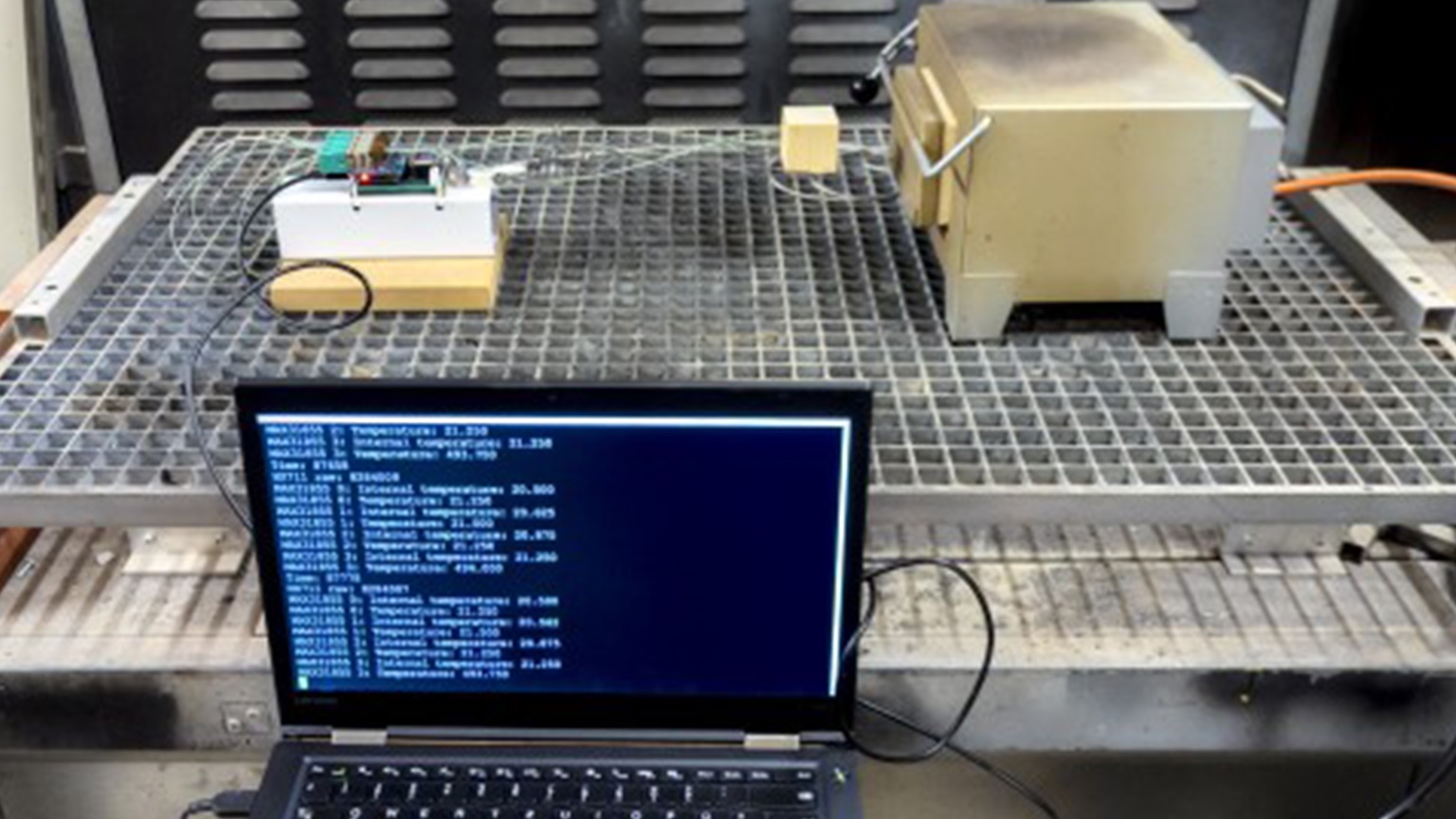}\hfill{}
\includegraphics[width=0.235\textwidth] {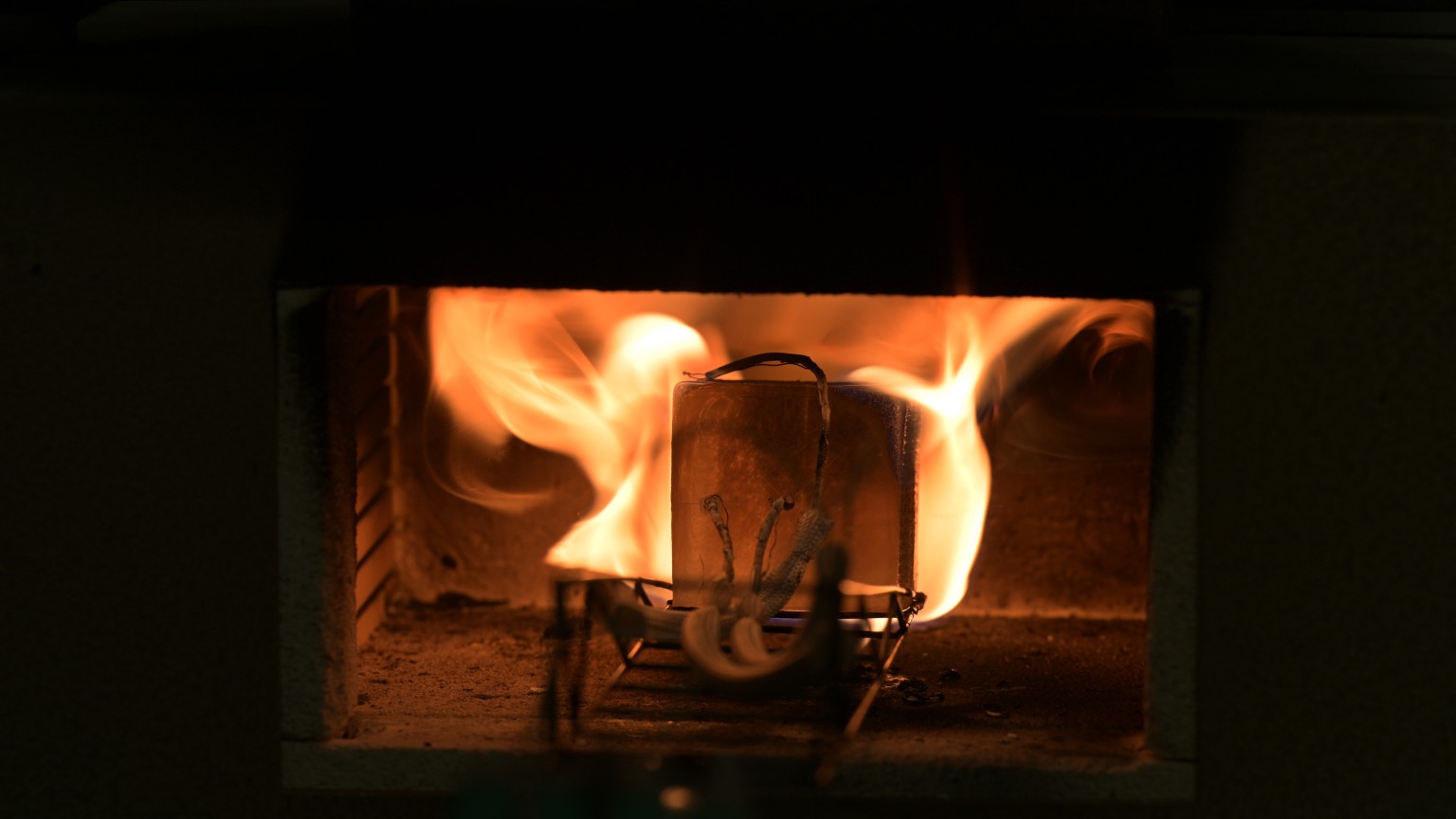}\hfill{}
\includegraphics[width=0.235\textwidth] {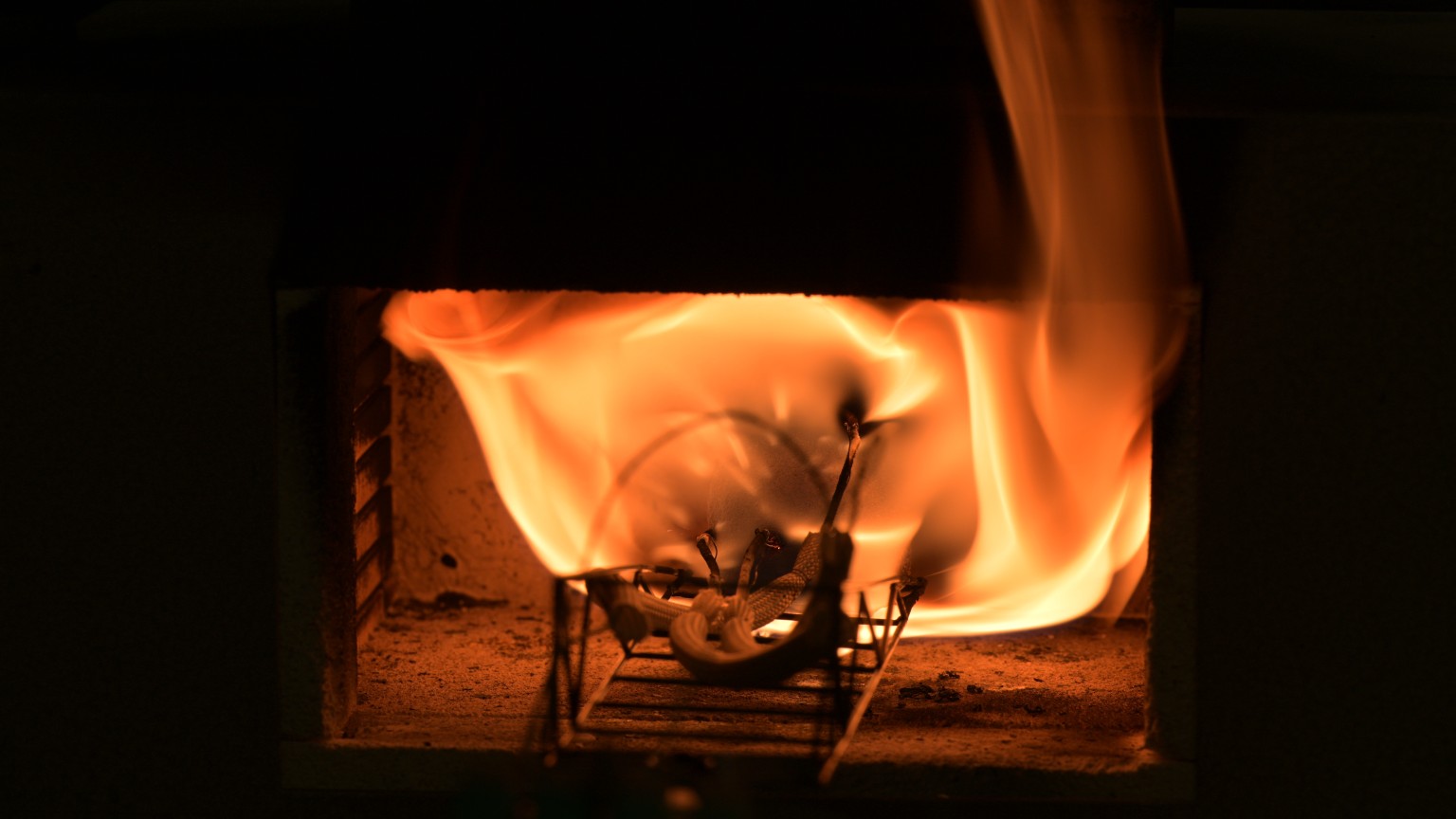}\hfill{}
\includegraphics[width=0.235\textwidth] {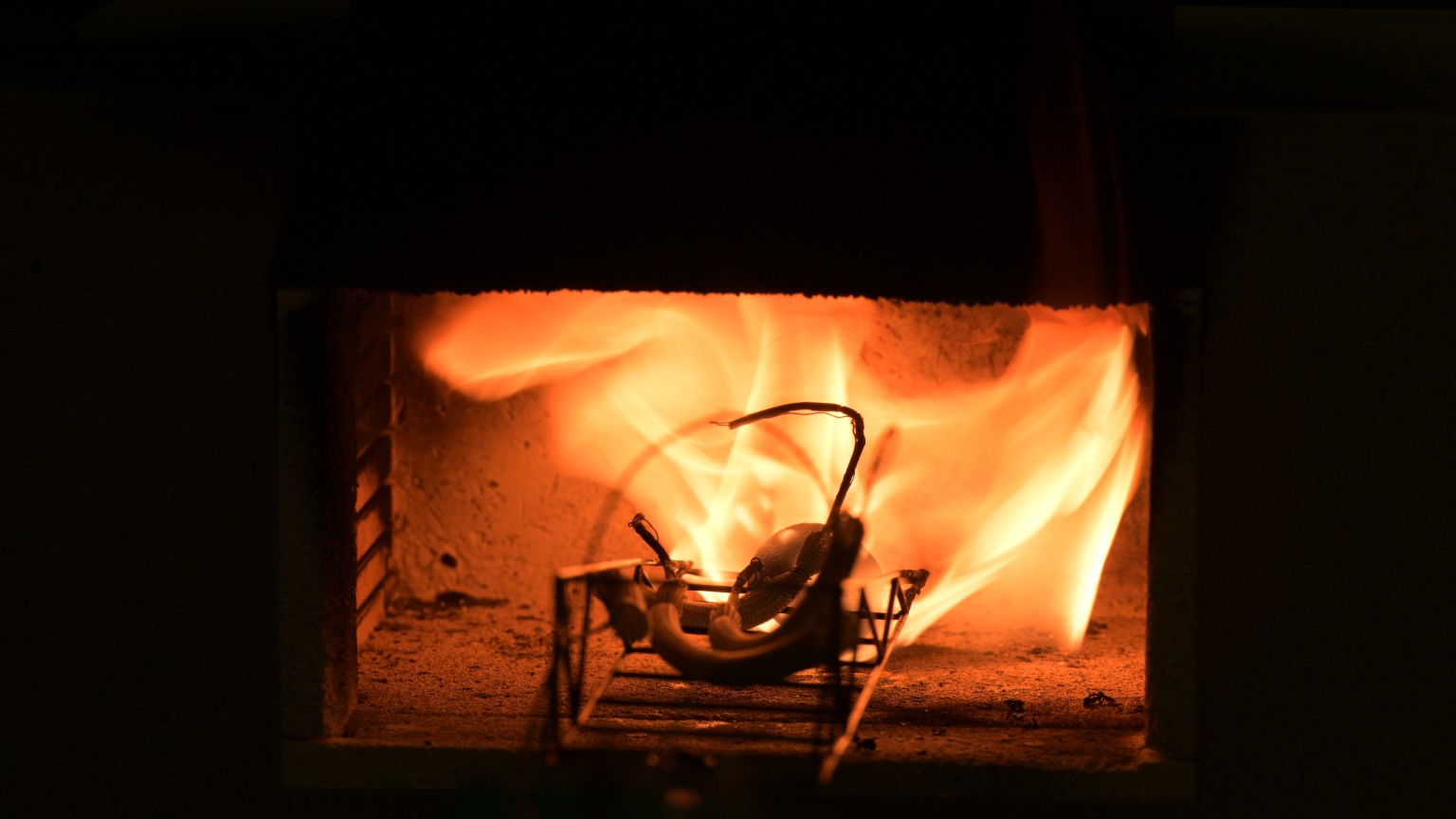} \hfill{} \rotatebox[x=0em, y=5em]{-90}{PMMA}\\
\hfill{}\includegraphics[width=0.754\textwidth]{figures/timeline.pdf}\\
\vspace{-0.4cm}
\caption{\label{fig:experiment_setup}Experimental setup for mass loss and temperature measurement.
Samples of different materials (wood and PMMA) are equipped with thermo probes and placed on a scale arm inside an oven that heats the sample up until the ignition temperature.
During the combustion process, the weight and temperature and different depths (surface, sub-surface, and core) are continuously recorded.}
\end{figure*}

To quantitatively assess the accuracy of our simulation, we performed a combustion experiment with different materials, shown in Figure~\ref{fig:experiment_setup}.
Our main motivation is to study volumetric temperature profiles inside the material as well as determining characteristic differences between charring and non-charring combustibles.

We take cubic blocks ($40\times40\times40$~mm$^3$) of different materials and drill holes into them in which \textit{Type K thermocouples} are installed for temperature measurements inside the volume during the combustion process at 20~mm and 5~mm depths.
A third thermocouple is placed right on the surface.
We placed the prepared samples on a scale inside an oven that increases the ambient temperature of the scene to ensure complete combustion of the samples.

\begin{figure*}
\includegraphics[width=0.32\textwidth]{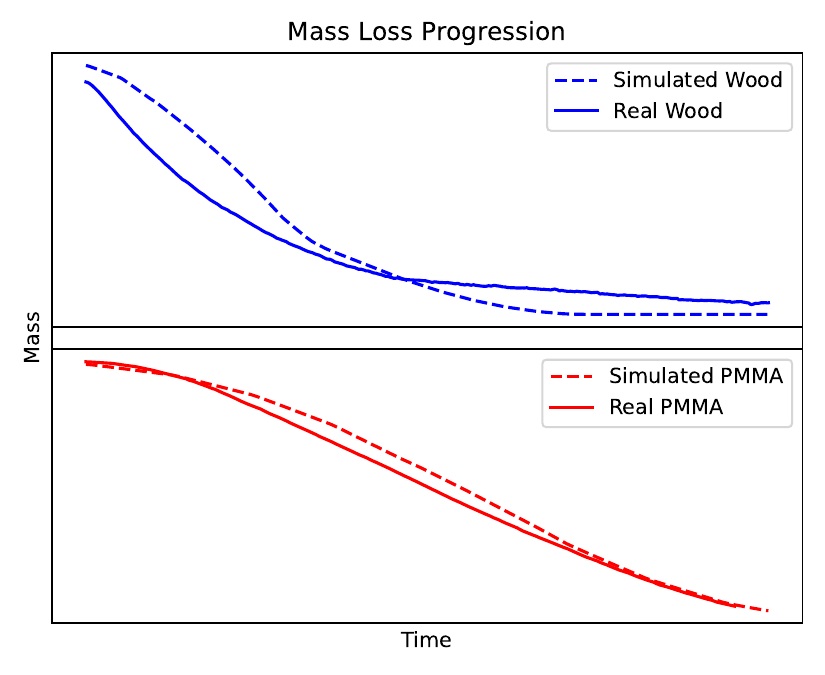}\hfill{}
\includegraphics[width=0.32\textwidth]{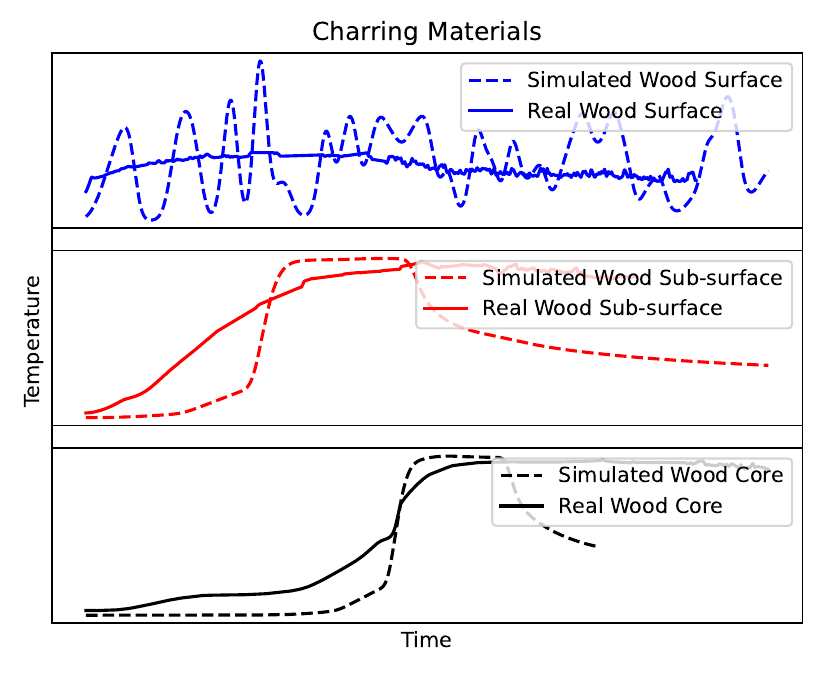}\hfill{}
\includegraphics[width=0.32\textwidth]{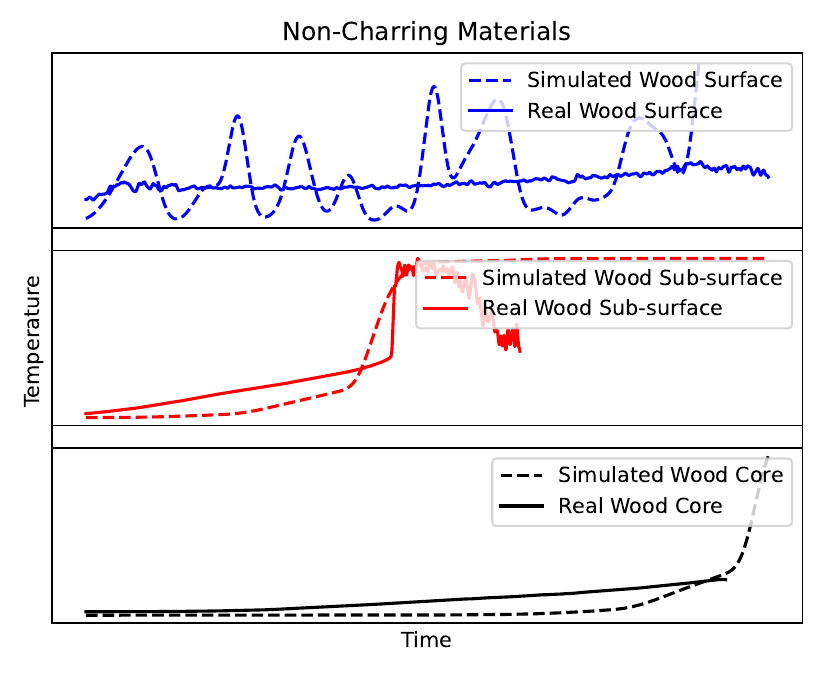}\\
\vspace{-0.3cm}
\caption{\label{fig:experiment_curves} \new{Comparisons of the temporal evolutions of mass and temperature between simulation and real experiment.
Left: Simulation
of a burning block with charring enabled and disabled (shown at the
same time step).
Middle: Temperature
curves for different simulation settings and measurements from the real
experiment of charring materials.
Right: Temperature
curves for different simulation settings and measurements from the real
experiment of non-charring materials.
Please note that due to the different materials being used, the horizontal axis is normalized to cover the full combustion process.}
}
\label{fig:experiment}
\end{figure*}

We then recreate a similar environment in our simulator and likewise record the sample's mass and temperature at identical positions.
The results are shown in Figure~\ref{fig:experiment_curves}.

For the mass curves, we find that charring materials show a parabola shaped curve while non-charring materials show a slightly S-shaped curve.

For the temperature curves, we expect that sensor readings closer to the surface rise sooner.
In the real experiment, we face however the problem that the thermocouples on and close to the surface cease to measure meaningful values as soon as they become detached from the sample due to combustion of the materials (as can be seen in the video).
This is a less severe problem for charring materials since the remaining char can hold the sensor in place longer.

Non-charring materials do not form an insulating layer.
Therefore, the temperature at the inner sensor rises rapidly, as the pyrolysis front approaches.
In the real experiment, the polymethyl methacrylate (PMMA) block dismantles shortly before the end of the experiment (see video), which is why the jump of the real PMMA core sensor is lacking.

For wood samples, the char layer acts as an insulator and generally smoothes out the curves.
We observe a distinct feature in both the real and simulated curves, where the sub-surface sensor curve has a more convex shape, while the core sensor curve is more concave.
This effect is likely caused by the char layers efficiency to shield the sample core from heat.

In conclusion, we report that our \textsc{FlameForge} simulator successfully reproduces the general trends of the measured curves providing the basis for accurately capturing combustion phenomena.
Please note that the exact calibration of our system to a given real environment is a very challenging task which we could partially address as part of this contribution.
Understandably, the slow rise of the inner sensors before their sudden rise in the non-charring case, is not found in the simulation.

\subsection{Varying Starting Positions}
\begin{figure*}
\centering
\includegraphics[width=0.11\textwidth]    {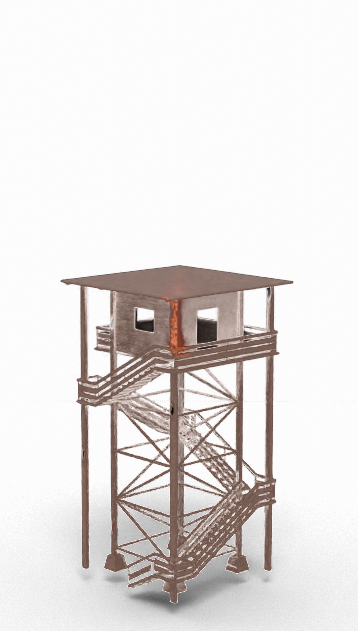}
\includegraphics[width=0.11\textwidth]    {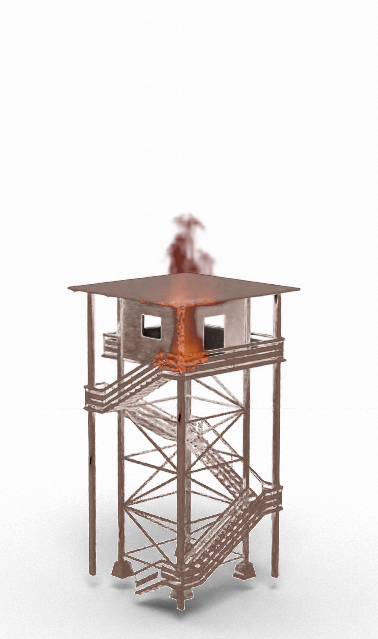}
\includegraphics[width=0.11\textwidth]    {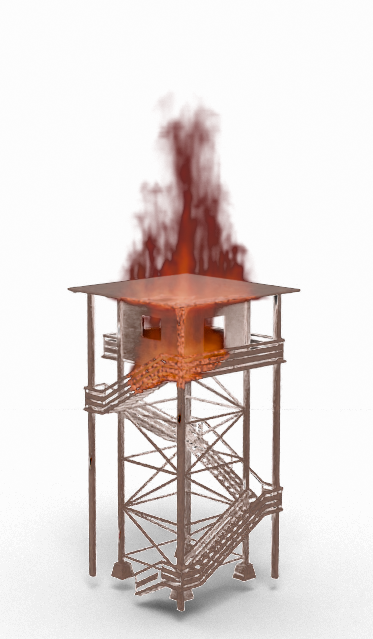}
\includegraphics[width=0.11\textwidth]    {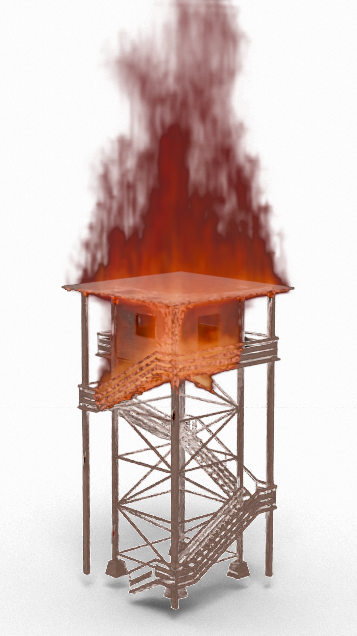}
\includegraphics[width=0.11\textwidth]    {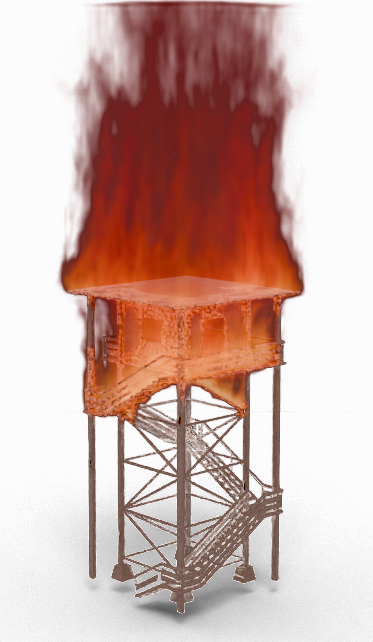}
\includegraphics[width=0.11\textwidth]    {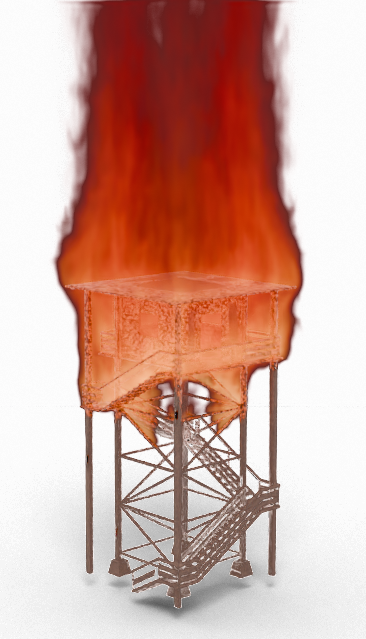}
\includegraphics[width=0.11\textwidth]   {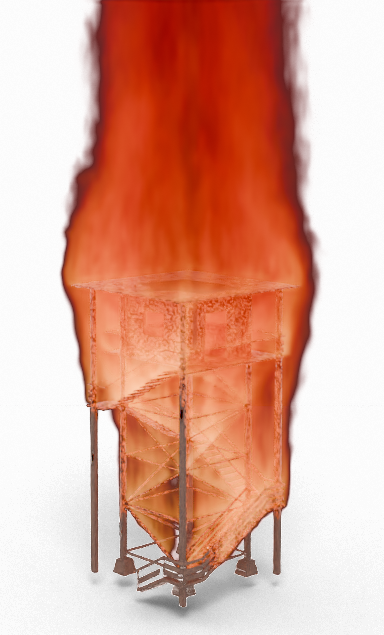}
\includegraphics[width=0.11\textwidth]   {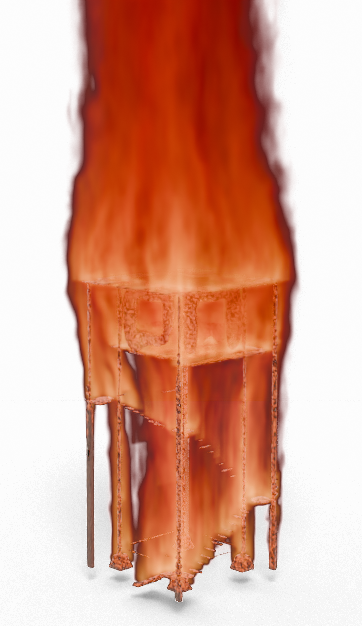} \\
\includegraphics[width=0.11\textwidth]   {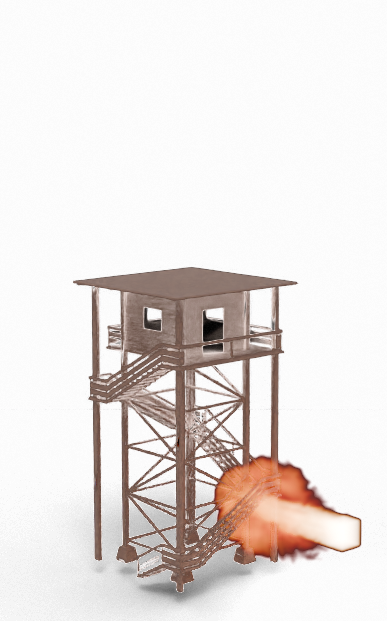}
\includegraphics[width=0.11\textwidth]   {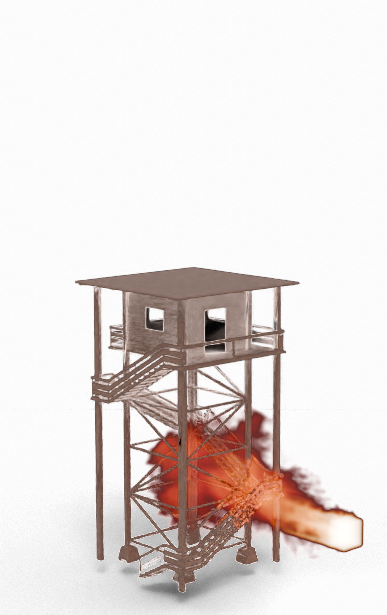}
\includegraphics[width=0.11\textwidth]   {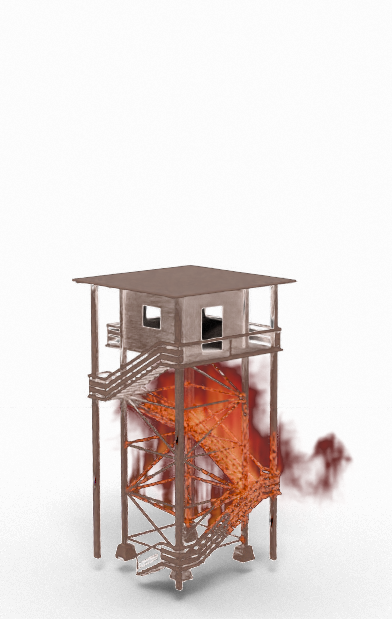}
\includegraphics[width=0.11\textwidth]   {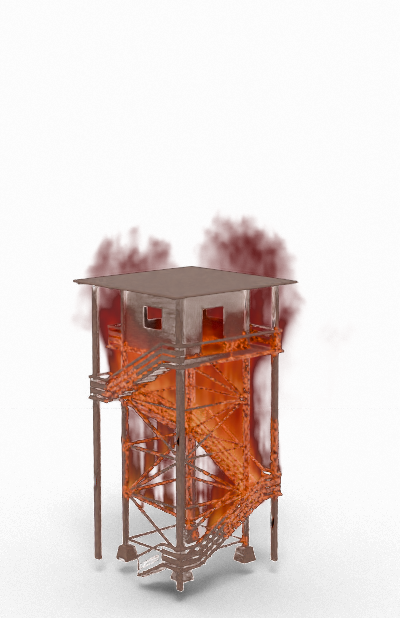}
\includegraphics[width=0.11\textwidth]   {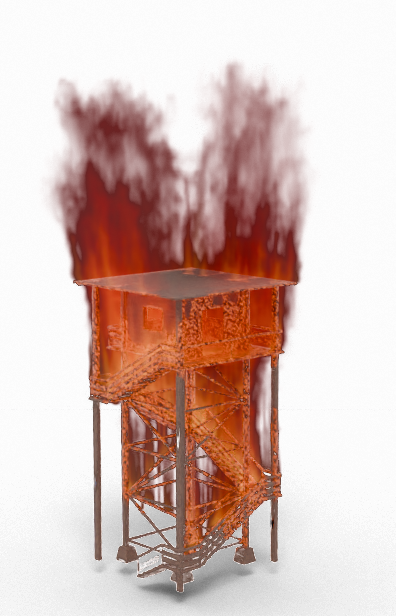} 
\includegraphics[width=0.11\textwidth]   {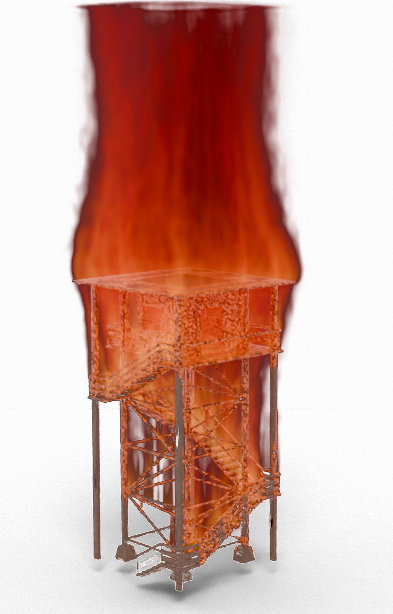} 
\includegraphics[width=0.11\textwidth]   {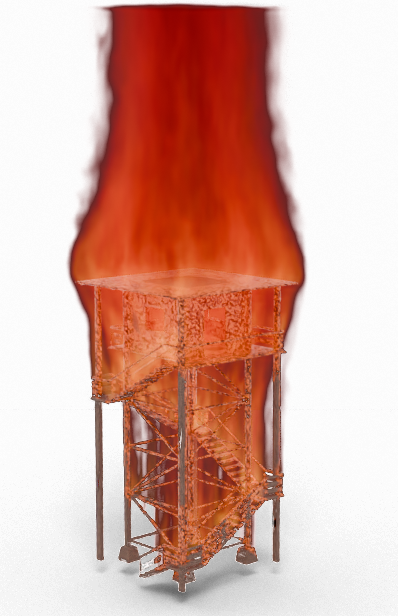} 
\includegraphics[width=0.11\textwidth]   {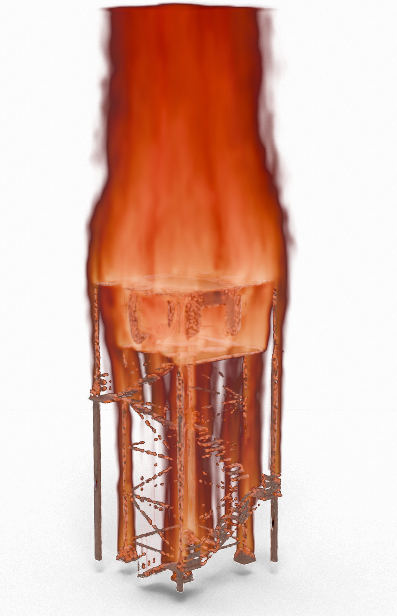} 
\\
\includegraphics[width=0.11\textwidth]   {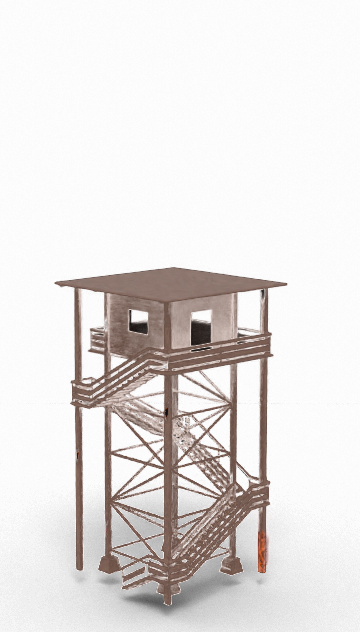}
\includegraphics[width=0.11\textwidth]   {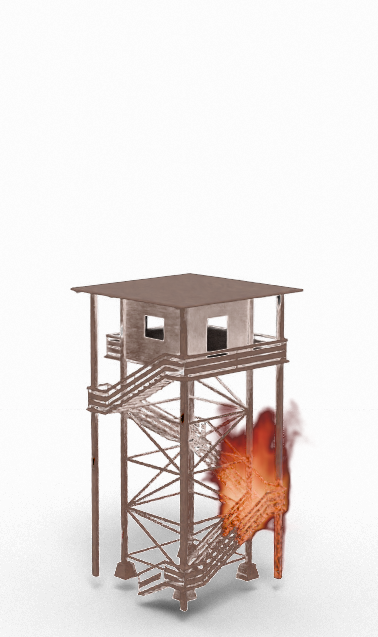}
\includegraphics[width=0.11\textwidth] {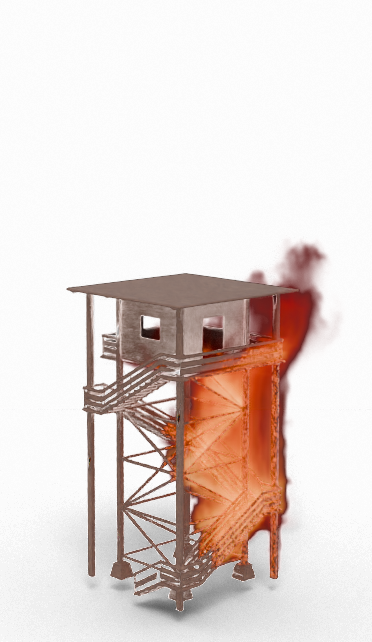}
\includegraphics[width=0.11\textwidth] {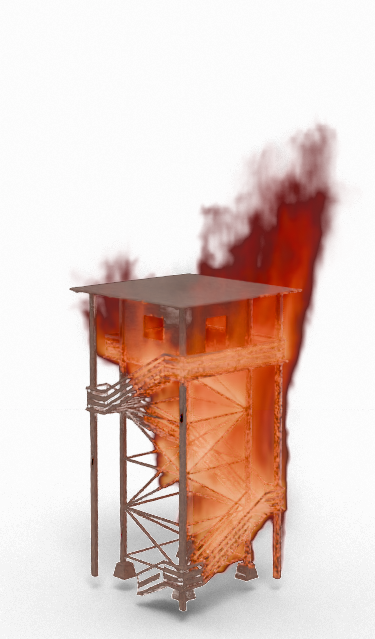}
\includegraphics[width=0.11\textwidth] {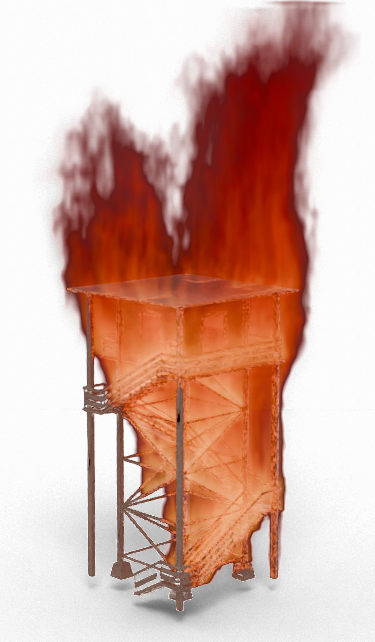}
\includegraphics[width=0.11\textwidth] {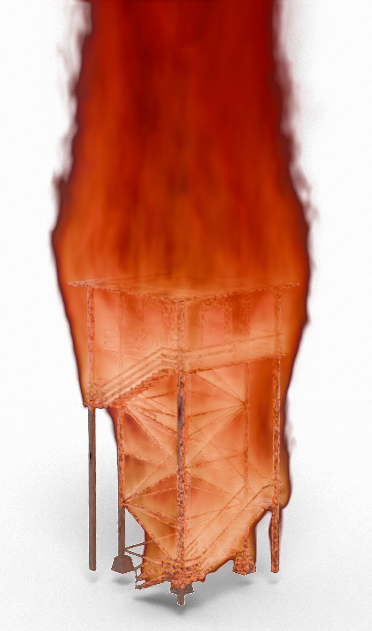}
\includegraphics[width=0.11\textwidth]{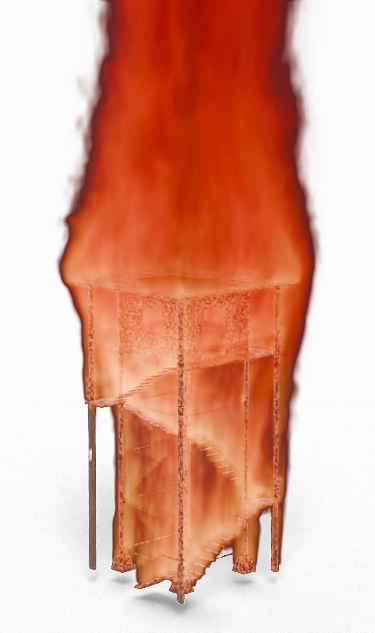}
\includegraphics[width=0.11\textwidth]{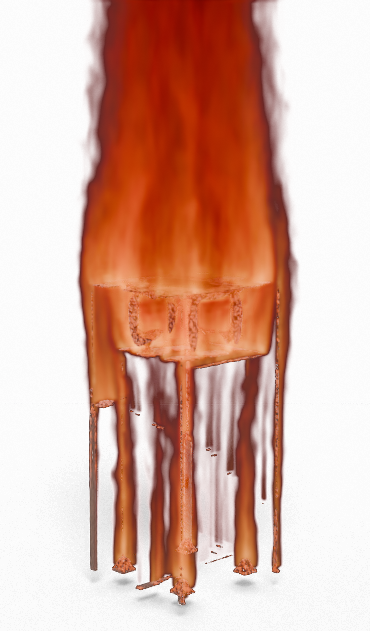} \\
\includegraphics[width=0.88\textwidth]{figures/timeline.pdf}\\
\vspace{-0.35cm}
\caption{\label{fig:result_tower}Simulation of a burning watchtower with different ignition locations. Top row: The fire starts at the roof. Middle row: The tower is ignited from the side using a flamethrower. Bottom row: The fire starts at the bottom of the tower.}
\end{figure*}

The same building can burn in drastically different ways depending on how the fire started.
We demonstrate this by burning a watchtower down in three different ways (Figure~\ref{fig:result_tower}):
A fire accidentally started near the main cabin on top (such as through a smoking incident), a massive burst from a flame thrower at its center, and a small fire started at its very bottom by an arsonist.
As expected, the flame thrower attack rapidly sets the whole tower ablaze.
The other two fires, though similarly small in the beginning develop very differently.
Due to buoyancy, fire has a strong preference to spread upwards, causing the fire at the bottom to quickly spread across the whole tower while the fire that started on top could have potentially been put out in time.

\subsection{Environmental Conditions and Multi-Object Scenes}
\begin{figure*}
\includegraphics[width=0.245\textwidth] {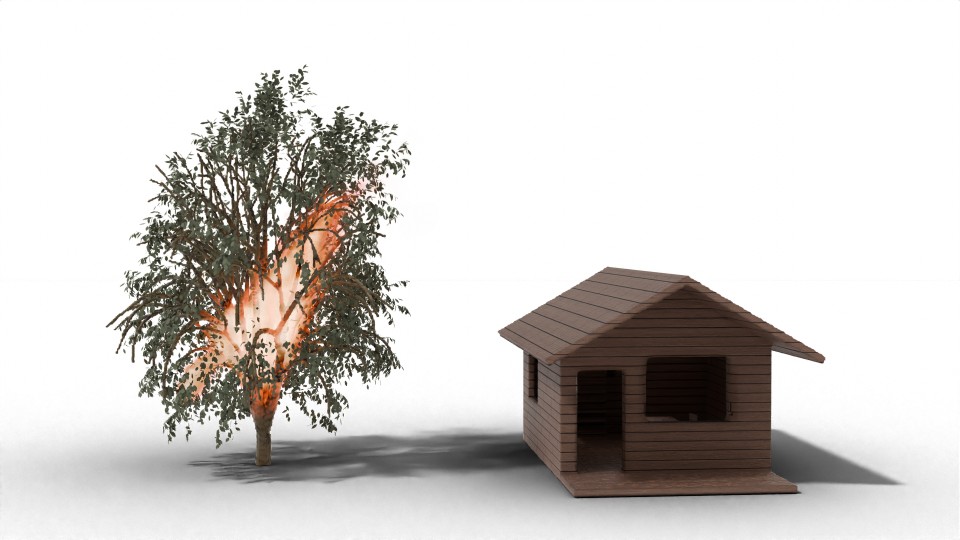}\hfill{}
\includegraphics[width=0.245\textwidth] {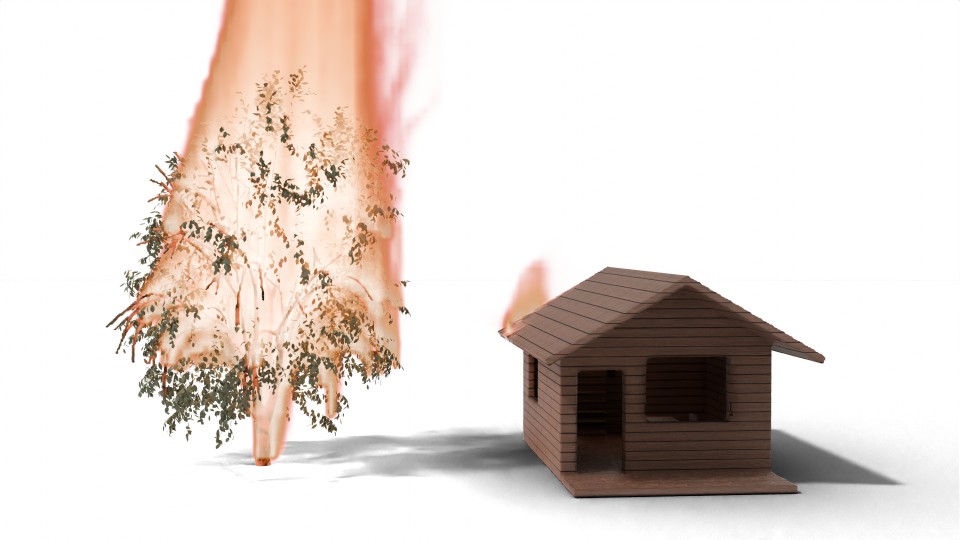}\hfill{}
\includegraphics[width=0.245\textwidth] {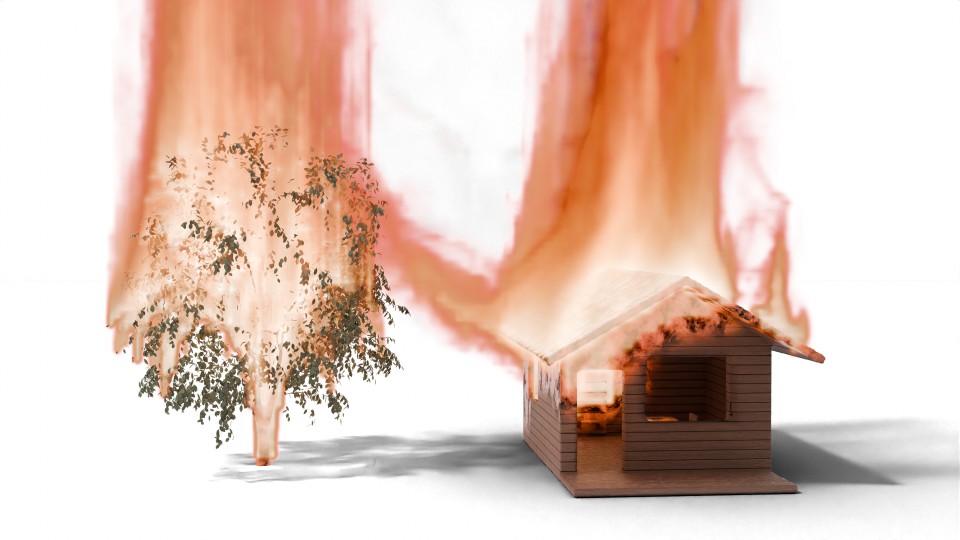}\hfill{}
\includegraphics[width=0.245\textwidth] {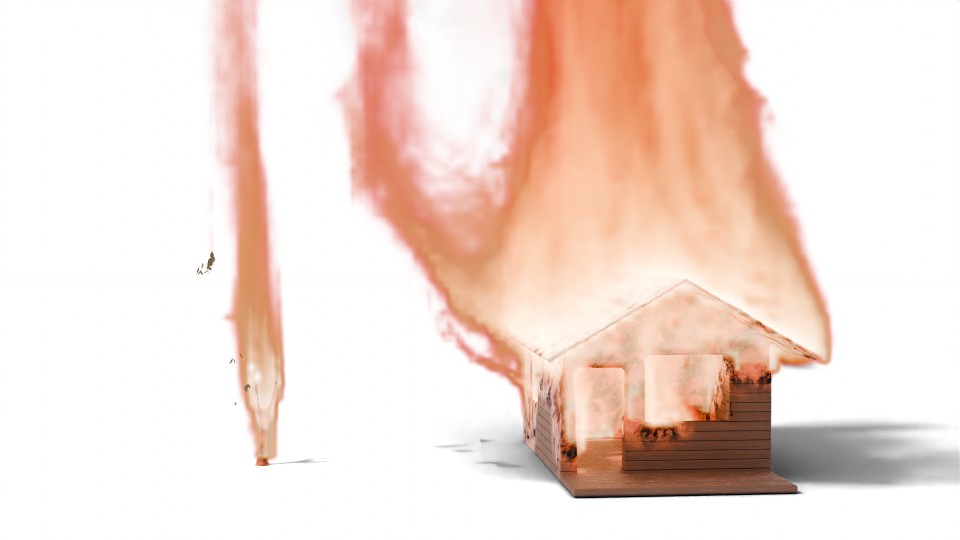}\\
\includegraphics[width=0.245\textwidth] {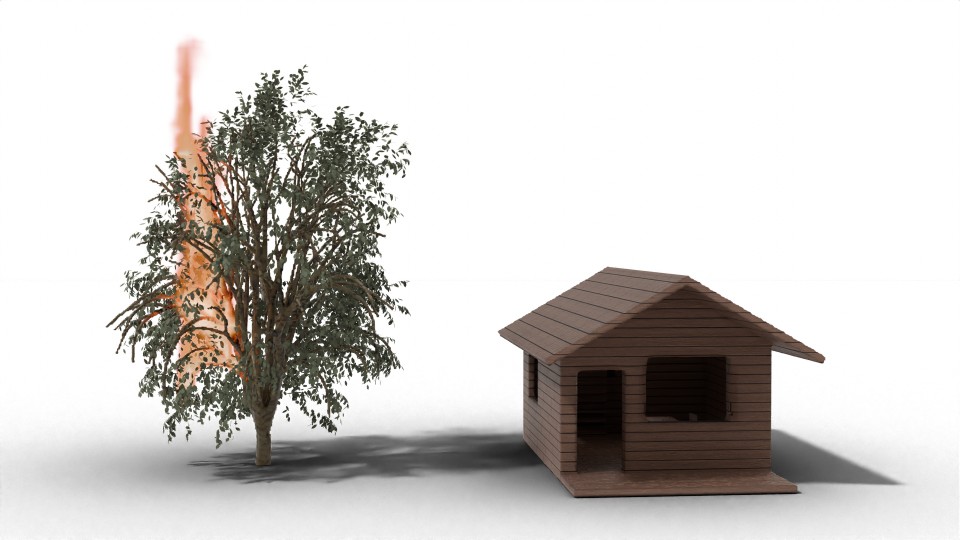}\hfill{}
\includegraphics[width=0.245\textwidth] {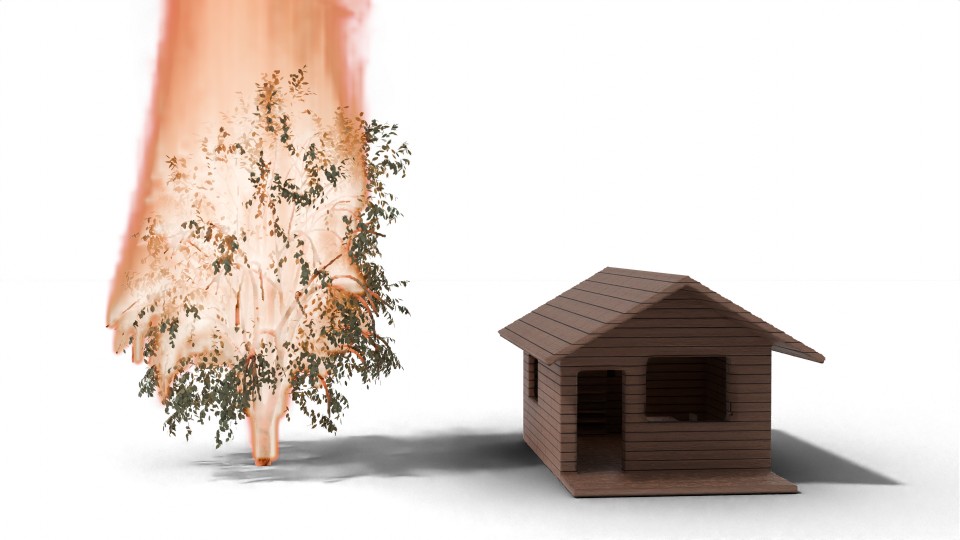}\hfill{}
\includegraphics[width=0.245\textwidth] {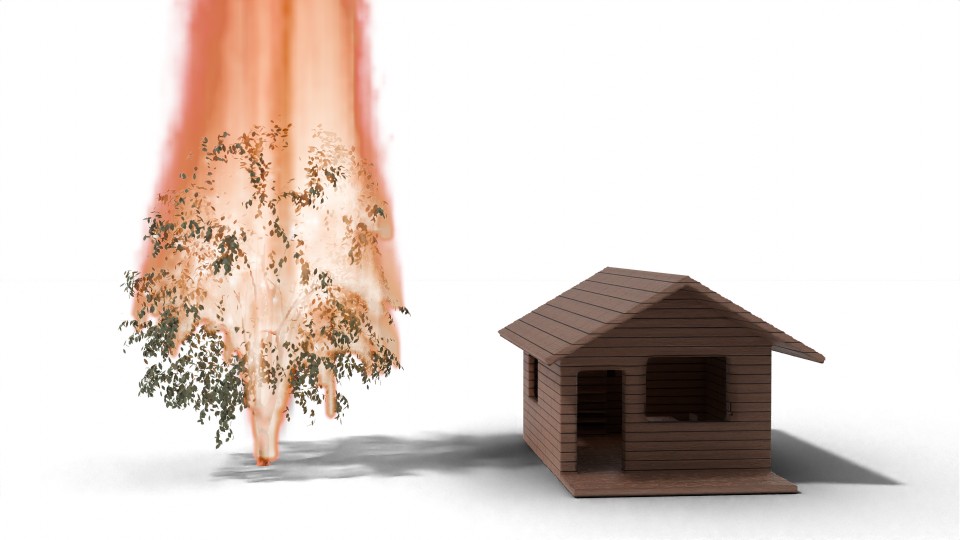}\hfill{}
\includegraphics[width=0.245\textwidth] {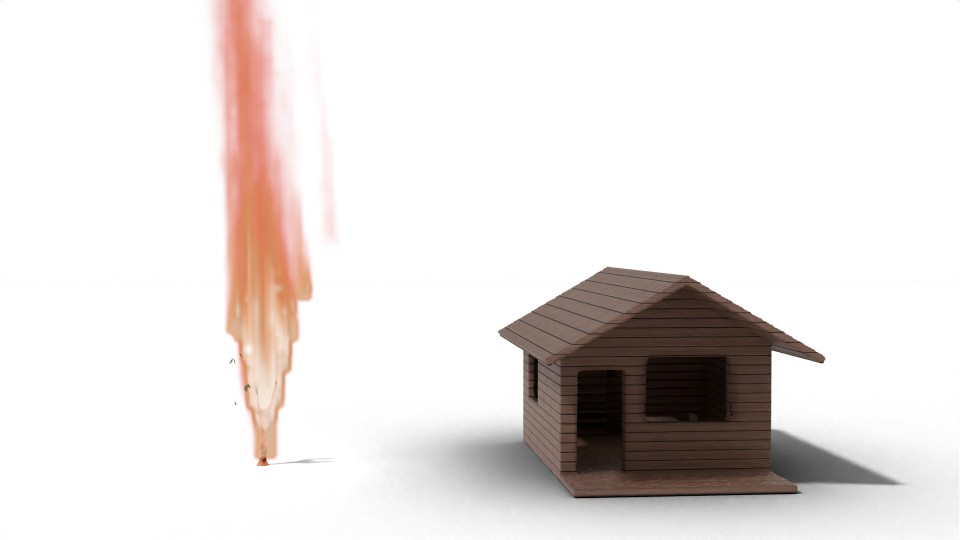}\\
\includegraphics[width=1.0\textwidth]{figures/timeline.pdf}\\
\vspace{0.15cm}
\hfill{} $t_0$ \hfill{} \hfill{} $t_1$ \hfill{} \hfill{} $t_2$ \hfill{} \hfill{} $t_3$ \hfill{}
\vspace{-0.2cm}
\caption{\label{fig:result_tree_house}Simulation of fire spread across multiple objects: A tree is on fire ($t_0$) which further spreads according to the wind direction igniting the roof ($t_1$) of a house (top row) causing severe damage ($t_2$). In contrast, if almost no wind is present, also in the long term ($t_3$), the house remains untouched (bottom row).}
\end{figure*}

In Figure~\ref{fig:result_tree_house} we show a scene consisting of multiple objects.
They each occupy their own local grid but are linked through the heat transfer of the surrounding air.
The velocity field $\textbf{u}$ can be externally excited to create wind.
Two versions of the same scene are shown:
With wind enabled, sufficient heat travels from the tree to the house and finally ignites the roof, causing the whole building to burn down.
Otherwise, just the tree burns down, leaving the house untouched.

\subsection{\new{Fracture Mechanics}}

Fracture mechanics plays a crucial role in the accuracy and reliability of combustion simulations. Therefore, appropriate models for mechanical simulations should be integrated into our simulator. Nevertheless, we deem this topic to be outside the scope of this paper as we are specifically focusing on the fire and combustion simulation. However, we exemplify the potential for two-way coupling between our simulator and additional mechanical simulators by including an example of a wooden bridge as shown in Figure~\ref{fig:burning-bridge}. Technical details are provided in Appendix~\ref{sec:appendix}.
\section{Conclusion}

Throughout this paper, we have examined the numerical simulation of the different aspects of wood combustion, considering factors such as material composition and structural configurations.
Our quantitative experiments show that the proposed mathematical model is capable of recreating non-trivial aspects of volumetric heat transfer during combustion processes that ultimately govern the large scale spread of fire. However, a number of extensions should be explored in future work.

The gaseous component of \textsc{FlameForge} is simplified by being represented only as air with temperature; future work could incorporate other chemicals involved in the combustion process.
For instance, the speed of the combustion reaction critically depends on the availability of oxygen in the surrounding air.
Especially in enclosed areas, a lack of oxygen can therefore greatly alter the development of a fire.

Within the scope of this paper, we have exemplified the potential for two-way coupling of our \textsc{FlameForge} simulator for the combustion of generalized (static) wooden structures and dynamic rods simulated via position-based dynamics.
However, for practical use, a general fracture model for the individual solid parts of the different scenes should also be integrated.
Due to the constant change of the geometry caused by the consumption of matter, this is a challenging task to solve.
It would however greatly increase the plausibility of the destruction by fire.
%
Finally, to accelerate the simulation, our implementation currently using OpenVDB could be improved utilizing, e.g., NanoVDB~\cite{museth2021nanovdb}, or NeuralVDB~\cite{kim2022neuralvdb} for simulations at interactive rates.

\begin{figure*}
\includegraphics[width=0.33\textwidth]{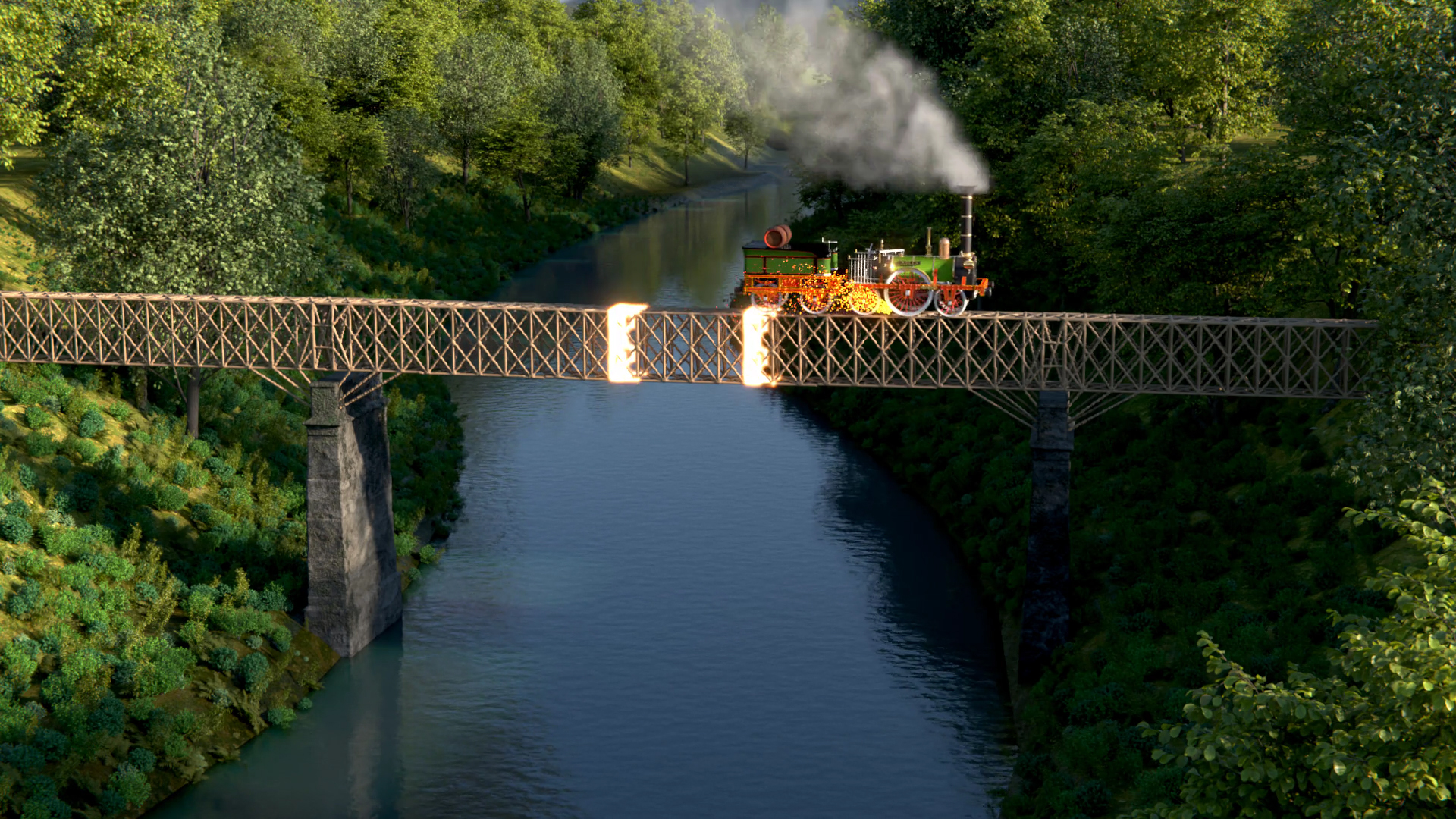}
\includegraphics[width=0.33\textwidth]{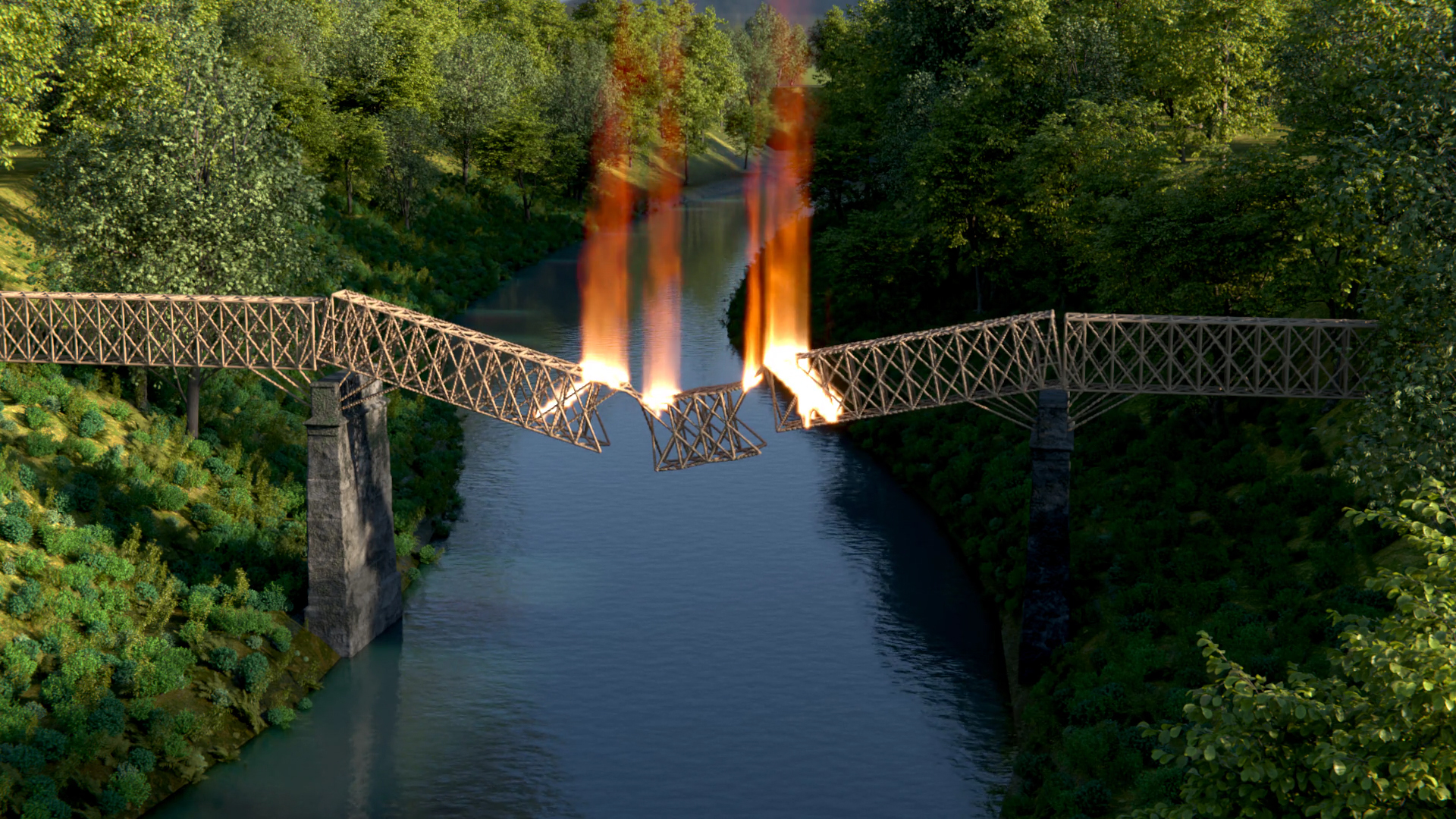}
\includegraphics[width=0.33\textwidth]{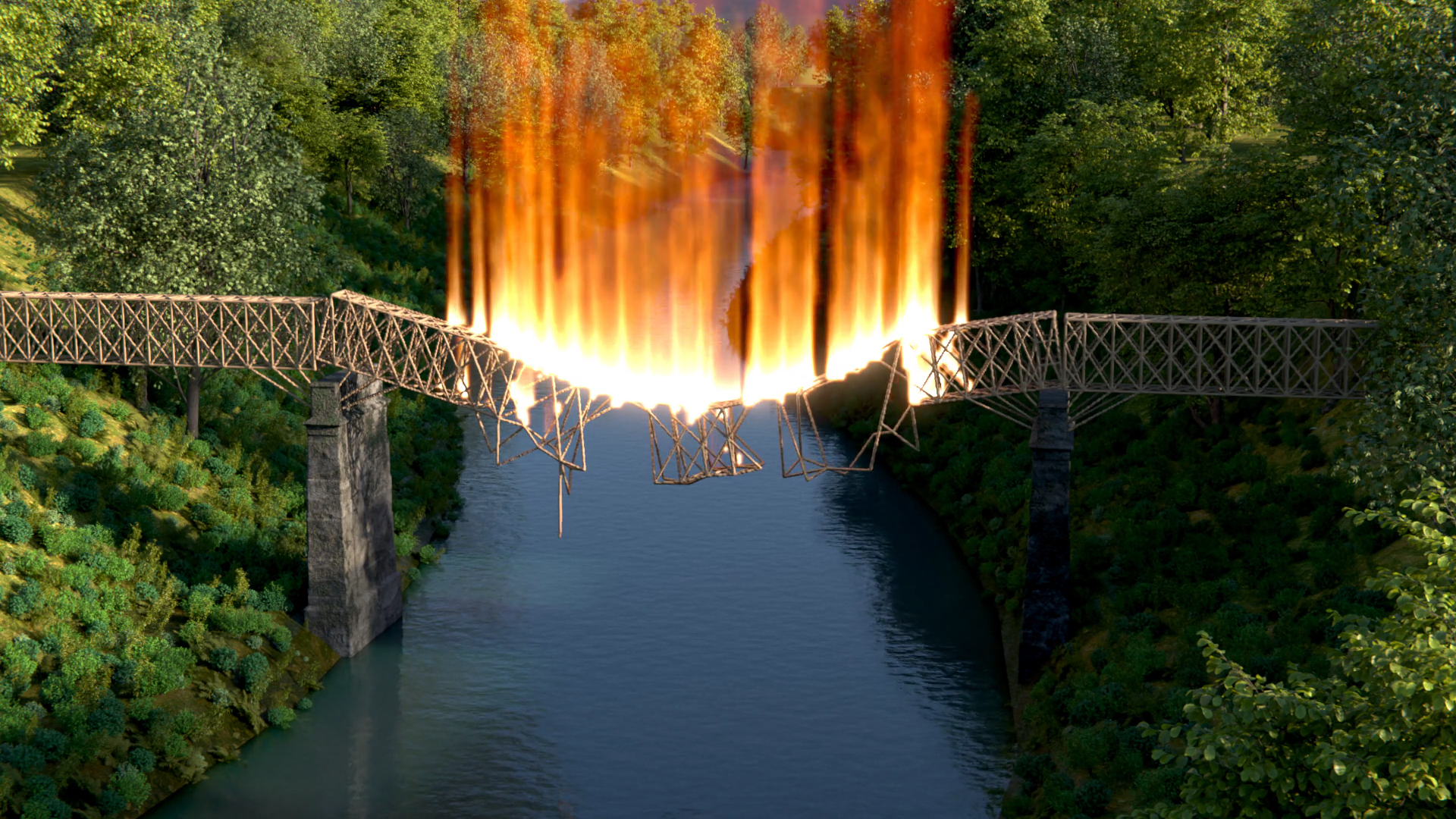}
\includegraphics[width=0.33\textwidth]{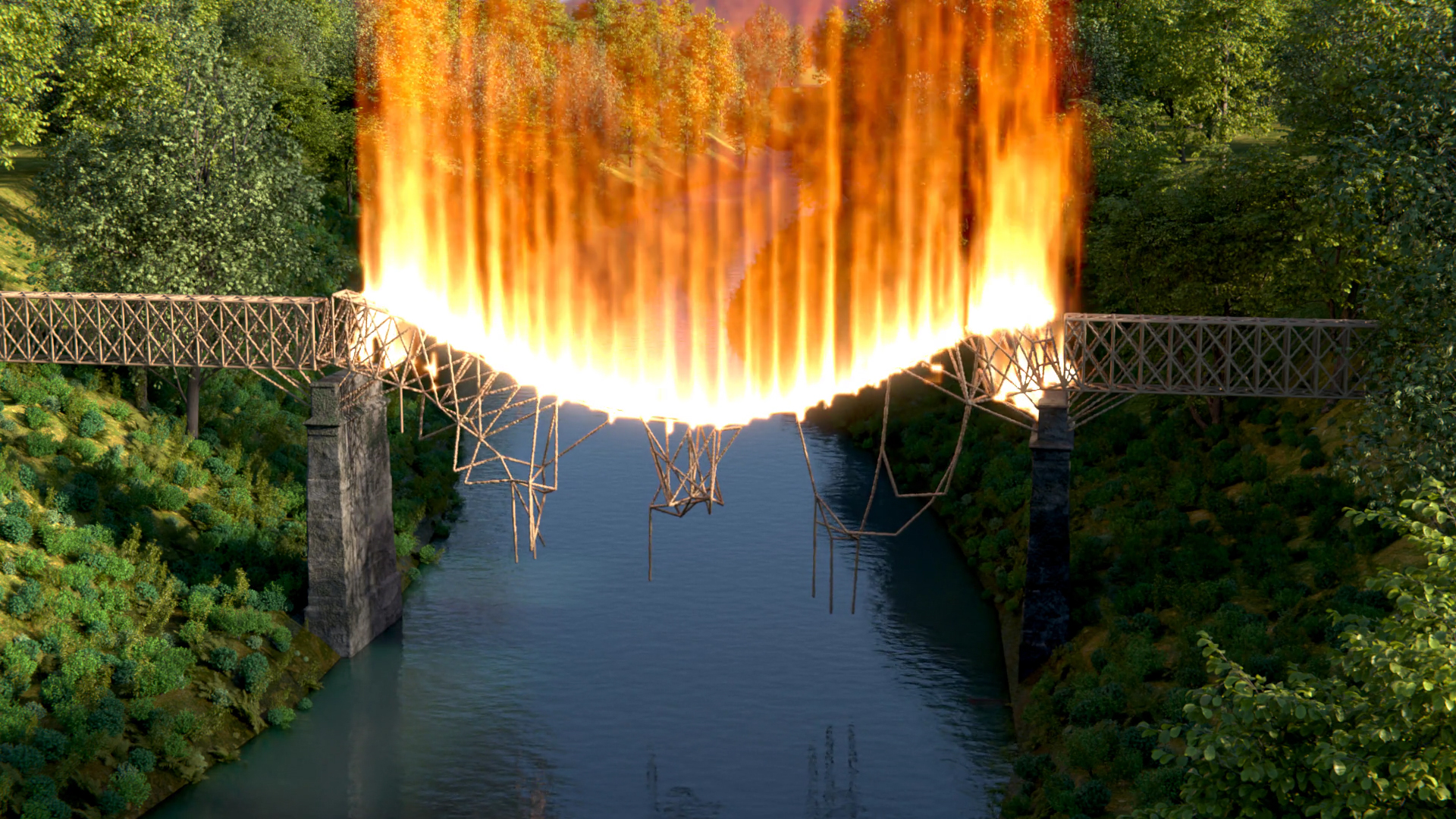}
\includegraphics[width=0.33\textwidth]{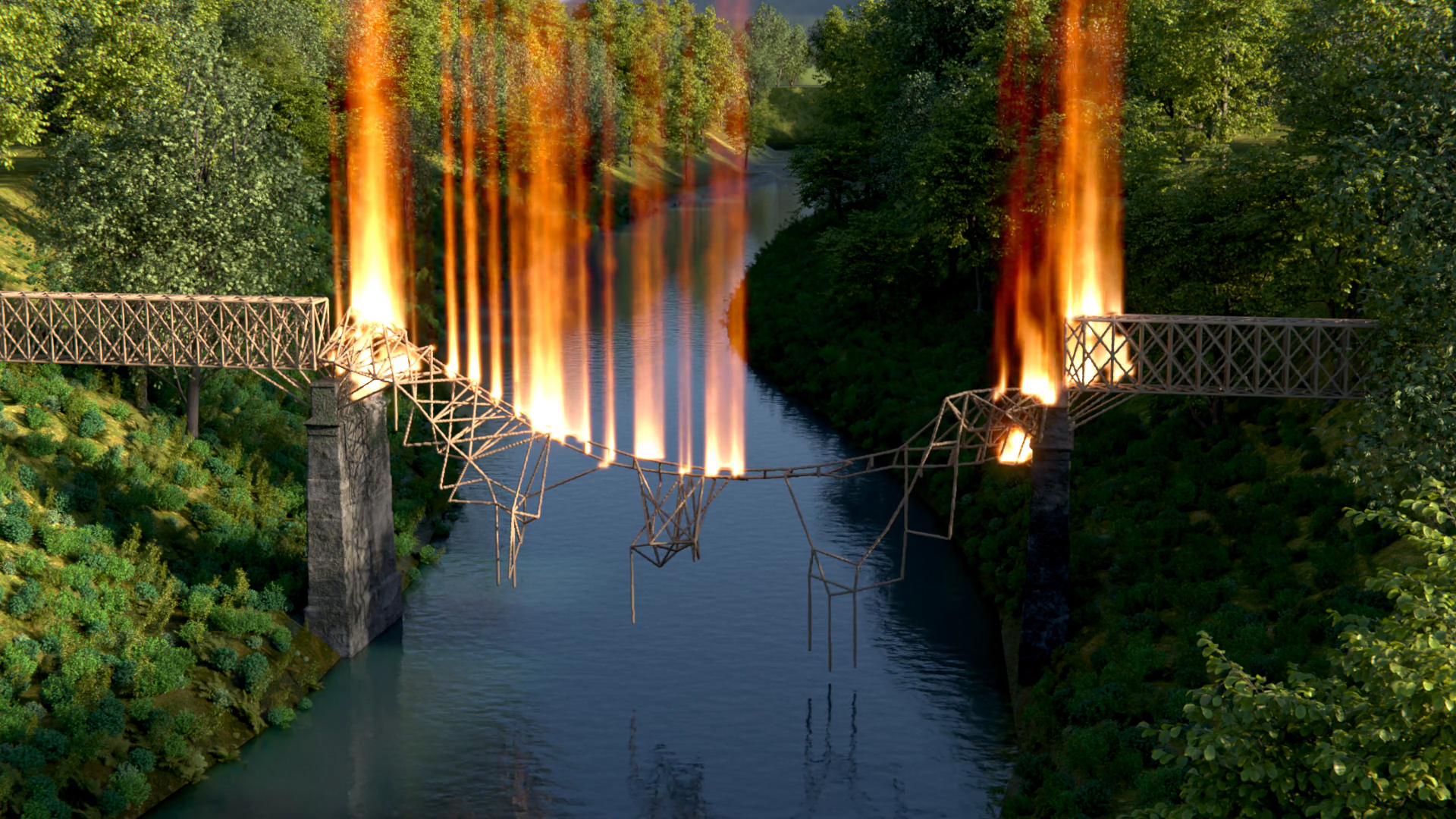}
\includegraphics[width=0.33\textwidth]{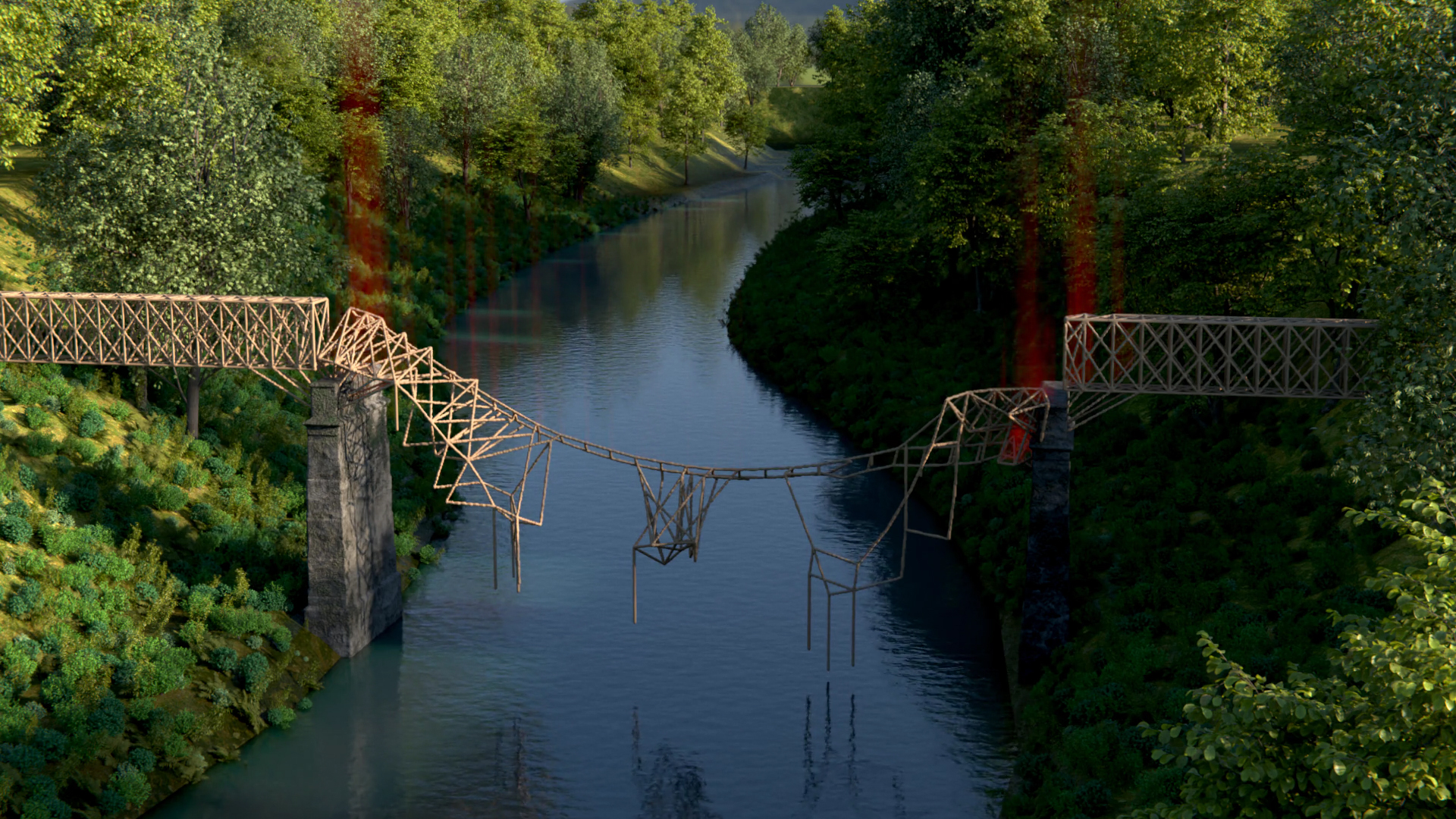}
\caption{\label{fig:burning-bridge}\new{The historical \textit{King Louis Bridge} built by Royal Bavarian State Railways (1847 to 1851) has been predominantly constructed from wood (larch, oak, and pine) as well as from iron and stone. This historical landmark of civil engineering -- the oldest surviving bridge erected by William Howe -- has originally been encased safeguarding the wood from glowing pieces of coal that might have fallen from locomotives for fire protection. We model such a scenario showcasing the capabilities of our \textsc{FlameForge} simulator. Glowing fragments fall onto railroad ties, sparking two fire sources (top row, left). The fire intensifies causing strong deformations and the characteristic sagging effect can be observed (top row, middle and right). The fire further intensifies (bottom row, left) until it can eventually be extinguished (bottom row, middle). As the fire is almost extinguished, the bridge's central element has been severely damaged while the tracks are finally drooping in the form of a catenoid (bottom row, right).}}
\end{figure*}

\bibliographystyle{eg-alpha} 
\bibliography{references}  

\section*{Acknowledgements}
This work has been partially funded by KAUST through the baseline funding of the Computational Sciences Group.

\appendix
\section{Fracture Mechanics}
\label{sec:appendix}
Fracture mechanics is essential for ensuring the accuracy and reliability of combustion simulations, necessitating the integration of suitable mechanical models into our simulator. However, this paper specifically focuses on fire and combustion simulation, so we do not delve into this aspect. Instead, we demonstrate the potential for two-way coupling between our simulator and additional mechanical simulators with an example involving a wooden bridge.

Specifically, we simulate a fire scenario at the historic wooden railway bridge spanning the Iller river in the town of Kempten, located in Allgovia, a region in Swabia, southern Germany. Although presently supported by a steel composite construction and used as a pedestrian and bicycle bridge, this historic railway bridge, named after King Ludwig I of Bavaria (\textit{King Louis Bridge}), was originally predominantly constructed from wood.\footnote{\new{\url{https://structurae.net/en/structures/king-louis-bridge}}} The bridge built by Royal Bavarian State Railways from 1847 to 1851 spanning the valley -- which is $120$~m wide and $34$~m deep at this point -- with three girders, has been recognized as a historical landmark of civil engineering by the Federal Chamber of Engineers in Germany in 2012. It is regarded as the oldest surviving bridge erected by the renowned American bridge builder William Howe. Originally, the bridge has been encased to shield the wood from weather elements, and likely also to safeguard it from glowing pieces of coal that might have fallen from locomotives for fire protection. We model such a scenario where glowing fragments fall onto railroad ties, sparking fires which cause severe damage. The results are shown in Figure~\ref{fig:teaser}.

\begin{figure}
\centering
\vspace{-0.5cm}
\includegraphics[width=1.0\columnwidth]{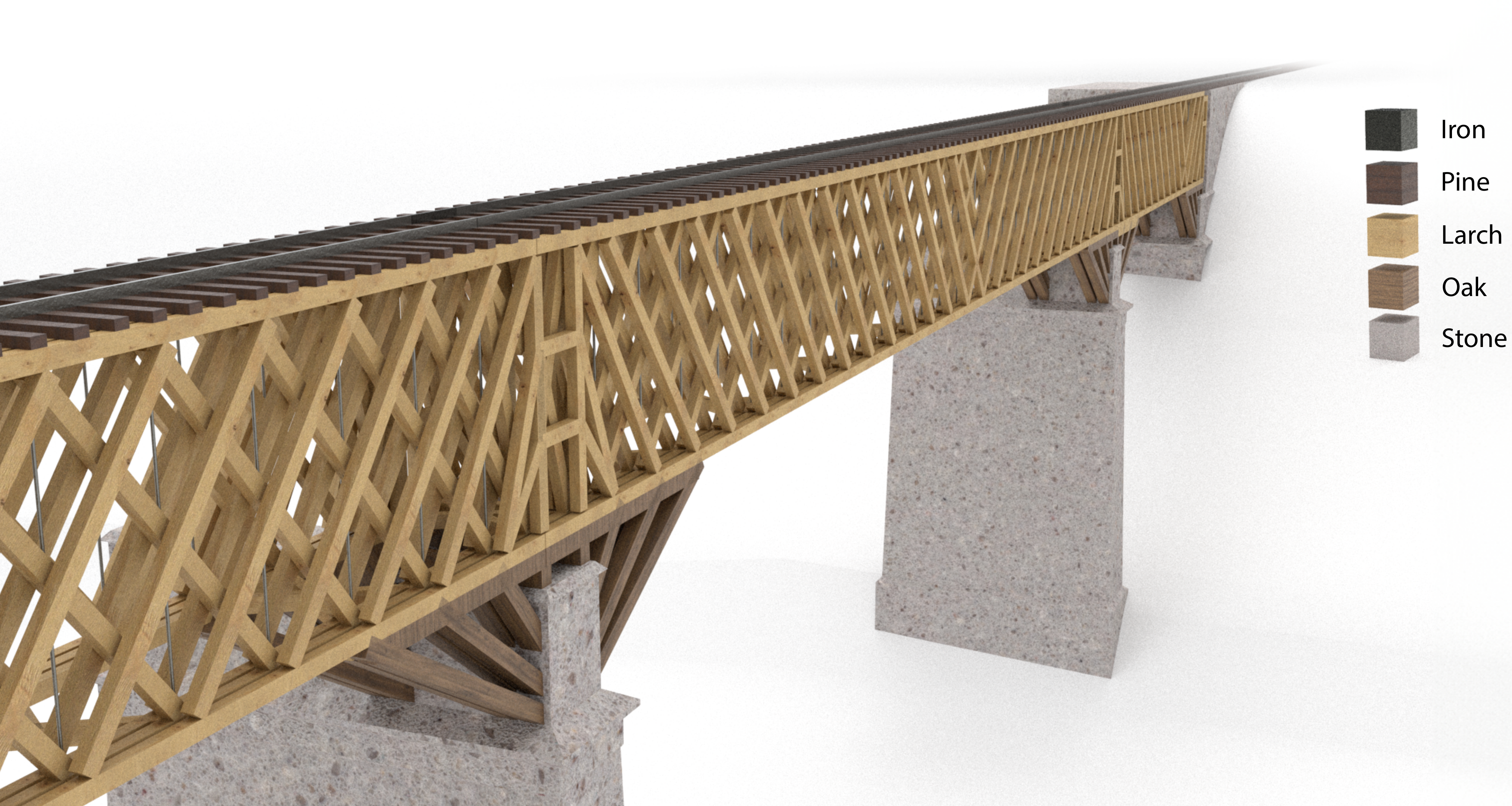}
\vspace{-0.15cm}
\caption{\label{fig:bridge}\new{Illustration of the main element of the \textit{King Louis Bridge}: Between the two stone pillars (light gray), the bridge is has been made almost entirely of wood (light brown for larch and dark brown for oak) as it is cheap and can absorb compressive forces well. However, wood does not tolerate tensile forces well, especially at the joints for which reason Howe used round iron rods (dark gray) for the vertical struts of the truss bridge, which were tensioned with screw nuts. As illustrated here, the diagonal compression struts were still made of wood. The railroad ties have been modeled from pine wood (wenge) and the tracks have been made out of iron (dark gray). The element shown here is $52$~m long.}}
\end{figure}

To accomplish this, we have modeled the bridge in \textit{Rhinoceros 3D}\footnote{\new{\url{https://www.rhino3d.com/}}} as illustrated in Figure~\ref{fig:bridge}. Three different types of wood have been used (larch, oak, and pine) as well as iron and stone. The corresponding parameters for material density and stiffness have been taken from the literature (see Blankenhorn~\shortcite{BLANKENHORN20019722} and references therein). We exported the bridge as a graph structure to simulate its mechanical behavior using position-based dynamics~\cite{10.2312:egt.20151045}, and as a density field for the combustion simulation.

The dynamic simulation using position-based dynamics is rod-based, while our fire and combustion simulation is voxel-based. To couple these simulations, we need a robust mapping between domains, which requires defining rasterizing and extraction operations.

For rasterizing, the mass density field is rebuilt from scratch using rod information, with indices saved to accurately separate masses of overlapping voxels. For extraction, the rods record the amount of mass they cover.

Only density is mapped to the dynamic simulation, while temperature is reused from the previous step.

Typically, a long time scale is simulated, with the dynamic simulation running until convergence. The limitation of this approach is that the dynamic movement is somewhat constrained by temperature propagation; otherwise, the fire would extinguish.

\newpage

\end{document}